\documentclass[twocolumn, nofootinbib, aps, prd, floats, floatfix, amsmath, amssymb, longbibliography, superscriptaddress, preprintnumbers]{revtex4-1} 
\pdfoutput=1
\usepackage[final]{graphicx}
\usepackage{hyperref}
\usepackage{amsmath}
\usepackage{bm}
\usepackage{amsfonts}
\usepackage{amssymb}
\usepackage{latexsym}
\usepackage{graphicx}
\usepackage[english]{babel}
\usepackage{multirow}
\usepackage{float}
\usepackage{url}
\usepackage{slashed}
\usepackage{xcolor} 
\usepackage[utf8]{inputenc}
\usepackage{stmaryrd} 
\usepackage{enumitem}
\usepackage{hyperref}
\usepackage{cleveref}
\usepackage{siunitx}
\usepackage{verbatim}
\usepackage{chngcntr}
\usepackage{amsthm}
\usepackage{ulem}
\usepackage[switch]{lineno}

\newcommand{\be}{\begin{equation}}
\newcommand{\ee}{\end{equation}}
\newcommand{\Eq}[1]{Eq.~(\ref{#1})}
\def\mL{\mathcal{L}}

\begin{document}

\title{
Unveiling Time-Varying Signals of Ultralight Bosonic Dark Matter at Collider and Beam Dump Experiments
}

\author{Jinhui Guo}
\email{guojh23@pku.edu.cn}
\affiliation{School of Physics and State Key Laboratory of Nuclear Physics and Technology, Peking University, Beijing 100871, China}

\author{Yuxuan He}
\email{heyx25@pku.edu.cn}
\affiliation{School of Physics and State Key Laboratory of Nuclear Physics and Technology, Peking University, Beijing 100871, China}

\author{Jia Liu}
\email{Corresponding author. jialiu@pku.edu.cn}
\affiliation{School of Physics and State Key Laboratory of Nuclear Physics and Technology, Peking University, Beijing 100871, China}
\affiliation{Center for High Energy Physics, Peking University, Beijing 100871, China}

\author{Xiao-Ping Wang}
\email{Corresponding author. hcwangxiaoping@buaa.edu.cn}
\affiliation{School of Physics, Beihang University, Beijing 100191, China}
\affiliation{Beijing Key Laboratory of Advanced Nuclear Materials and Physics, Beihang University, Beijing 100191, China}

\author{Ke-Pan Xie}
\email{Corresponding author. kpxie@buaa.edu.cn}
\affiliation{School of Physics, Beihang University, Beijing 100191, China}

\begin{abstract}
Abstract: The ultralight boson represents a promising dark matter candidate exhibiting unique wave-like behaviors. These properties could transfer to the dark mediator, such as the kinetic mixing dark photon, which can be a link between the dark and Standard Model sectors, resulting in periodic oscillations of its mass. We propose a method to detect ultralight dark matter using dark mediators in collider and beam dump experiments, distinguishing it from conventional atomic, molecular, and optical methods. The time-varying nature of dark mediator mass exhibits a double-peak spectrum, reducing traditional constraints by 1 to 2 orders of magnitude, due to decreased luminosity exposure in each resonant mass bin. To enhance sensitivity, we utilize event time-stamps in the CMS Open Data and demonstrate that this technique boosts sensitivity by approximately one order of magnitude compared to the time-blind method. Moreover, it proves effective in detecting the invisible decay of the dark mediator.
\end{abstract}
\maketitle

\section*{Introduction}

Ultralight bosons are a promising candidate for dark matter (DM) due to their wave-like behavior, which could alter DM activity at small scales. As a result, they have become a popular area of research~\cite{Ferreira:2020fam}. Recent proposals have suggested that ultralight DM could produce time-varying fundamental constants~\cite{Battaglieri:2017aum, Kimball2022Bibber, Antypas:2022asj}. These changes can be detected through atomic, molecular, and optical physics, the Oklo phenomenon, and astrophysical experiments~\cite{Uzan:2010pm, Safronova:2017xyt, Marsh:2021lqg}. Ultralight bosonic DM may also cause oscillations in Standard Model (SM) gauge couplings and fermion masses, where the oscillation period is related to the DM mass~\cite{Arvanitaki:2014faa, Stadnik:2014tta, Stadnik:2015kia}, as well as temporal changes due to topological defects~\cite{Pospelov:2012mt, Derevianko:2013oaa, Stadnik:2014cea}.

The kinetic mixing dark photon model~\cite{Okun:1982xi, Galison:1983pa, Holdom:1985ag} is an illustrative example that mediates the SM and the dark sector with a mixing strength $\epsilon$~\cite{Pospelov:2007mp, Arkani-Hamed:2008hhe, Essig:2013lka, Alexander:2016aln, Battaglieri:2017aum}. Particle experiments have placed strict limits on both $\epsilon$ and mass $m_{A'}$ \cite{Ilten:2018crw, Bauer:2018onh}. In particular, the parameter space that explains the recent muon $\left( g-2\right)_\mu$ excess~\cite{Bennett:2006fi, Muong-2:2021ojo} has been ruled out~\cite{Pospelov:2008zw, Bauer:2018onh}. However, if an ultralight scalar DM $\phi$ is charged under the dark $U(1)'$, it can induce periodic oscillations of the $A'$ mass, leading to a significant reduction in the existing collider and beam-dump bounds. This is especially true for the dilepton resonance searches~\cite{BaBar:2014zli, LHCb:2017trq, LHCb:2019vmc, LHCb:2020ysn, Merkel:2014avp, NA482:2015wmo, Duarte:2022feb, Katsuragawa:2022eug}, where even the previously highly excluded dark photon solution to muon $(g-2)_\mu$ can become viable again. Furthermore, we propose the use of event-by-event time stamps of recorded events and demonstrate with CMS Open Data, which could improve the sensitivity on the signal.

In this work, we investigate the implications of the time oscillation of the mediator mass in the dark sector on high energy colliders and beam dump experiments, instead of its direct coupling with the SM sector. This phenomenon has the potential to significantly alter the experimental constraints, as it generally reduces the luminosity exposure in each resonant mass bin. Additionally, the invariant mass spectrum exhibits a multi-peak (typically double-peak) feature, which is distinct from the traditional single-peak resonance.

\section*{Results}

\subsection*{Time-varying mass of the particle}

We assume the resonant particle has a time-dependent mass $m_{\rm res}(t)$ due to environmental effects and further take a time oscillating form with a period of $\tau$,
\begin{align}
	m_{\rm res}^2 (t)= m_{\rm res}^2 (t + \tau).
\end{align} 
For the resonant searches, the invariant mass of the event changes with time. Thus, the usual strategy of looking for resonance in a fixed bin suffers from reduced time exposure and leakage into other bins. If the data takes time $t_{\rm exp}\gg\tau$ and the experiment analyzes the full data in a time-blind way, then the relevant physical quantity is the time exposure $\Delta t_i$ in the $i$th mass bin $[ m_i, m_{i+1}]$, with the expression
\begin{align}
	\Delta t_i =\frac{t_{\rm exp}}{\tau} \int_{m_i}^{m_{i+1}} \left|  \frac{d t }{d  m_{\rm res}}  \right| d m_{\rm res}.
	\label{eq:time-interval}
\end{align}
Instead of a narrow resonance, the signal has a spread template fully determined by $ d t /d m_{\rm res}$. Then, the event number in $i$th bin is
\begin{align}
	N_i = \sigma_{\rm res}^{(i)} \epsilon_i \text{L} \frac{\Delta t_i}{t_{\rm exp}},
	\label{eq:signal strength}
\end{align}
where $\sigma_{\rm res}^{(i)}$ and $\epsilon_i$ are the resonant production cross-section and cut-efficiency for the $i$th bin, respectively, while \text{L} is the integrated luminosity. 
Since particle production and decay happen very quickly, a single event's resonant mass is unchanged. Therefore, the difference between our analysis and the previous resonant analysis is fully described by the time exposure fraction $\Delta t_i/t_{\rm exp}$.

There is a double-peak feature in the time-varying mass scenario that the peaks must show up at the minimum and maximum of resonant mass. Assuming the function $m_{\rm res}(t)$ is continuous and differentiable, the physical mass contains global minimum and maximum regardless of its periodic feature. Thus, it must have $d m_{\rm res}/dt = 0$ at two extreme points; therefore, the time exposure blows up accordingly. Additional local extrema can also contribute to peaks, leading to the multi-peak scenario. Other observable, if it is a function of time-varying mass, will inherit this property. 

\subsection*{Model setup}

We consider a kinetic mixing dark photon $A'$ with $U(1)'$ interaction,
\begin{align}
	\mathcal{L} = - \frac{1}{4} F'_{\mu \nu } F'^{\mu \nu} + \frac{1}{2}m_{0}^2 A'_\mu A'^{\mu} + \epsilon e A'_{\mu} J_{\rm em}^\mu,
\end{align}
where $\epsilon$ is the kinetic mixing strength which controls the strength of $A'$ coupling to electromagnetic current $J_{\rm em}$. The mass $m_0$ is a constant from  
$U(1)'$ spontaneously breaking. In addition, we consider a complex scalar DM $\phi$ with small charge $Q_\phi$ under $U(1)'$. The ultralight scalar DM obtains its relic abundance through misalignment mechanism \cite{Preskill:1982cy, Abbott:1982af, Dine:1982ah, Marsh:2015xka}, satisfying the equation of motion 
\begin{align}
	\ddot{\phi} + 3 H \dot{\phi} + m_\phi^2 \phi = 0 ,
	\label{eq: misalignment}
\end{align}
and at the late time, it is locally described by the classical wave function $\phi(t)$,
\begin{align}
	\phi(t) \approx \phi_1 \cos(m_\phi t) + \phi_2 \sin(m_\phi t)  ,
\end{align}
where $\phi_{1,2}$ are the complex field strengths, satisfying $\rho_{\rm DM} \approx \left(|\phi_1|^2 + |\phi_2|^2\right) m_\phi^2$ in the non-relativistic limit. In scalar quantum electrodynamics (QED), one can have the following four-point vertex 
\begin{align}
	\left(D_\mu \phi \right)^* D^\mu \phi \supset \left(g' Q_\phi \right)^2 \phi^* \phi A'_\mu A'^{\mu}.
\end{align}
It effectively leads to time oscillating $A'$ mass today as
\begin{align}
	& m_{A'}^2 (t)  = \tilde{m}_0^2 \left(1 + \kappa \cos^2\left( m_\phi t \right) \right), 
	\label{eq:oscillating-mass} \\
	& \tilde{m}_0^2 = m_0^2 + \left(g' Q_\phi \right)^2 \left( \phi_1^* \phi_1 + \phi_2^* \phi_2 - \sqrt{\xi^2 + \eta^2} \right), \\
	& \kappa  \equiv 2 (g' Q_\phi)^2 \sqrt{\xi^2 +\eta^2}/\tilde{m}_0^2,
\end{align} 
where $\kappa$ is the amplitude of the oscillation, $\xi = |\phi_1|^2 - |\phi_2|^2$ and $\eta = \phi_1 \phi_2^* + \phi_1^* \phi_2$. Thus, the oscillation mass is fully determined by three parameters, $\tilde{m}_0$, $\kappa$, and $m_\phi$, with the phase removed by the definition of $t=0$.
\begin{figure}
	\centering
	\includegraphics[width=0.48 \textwidth]{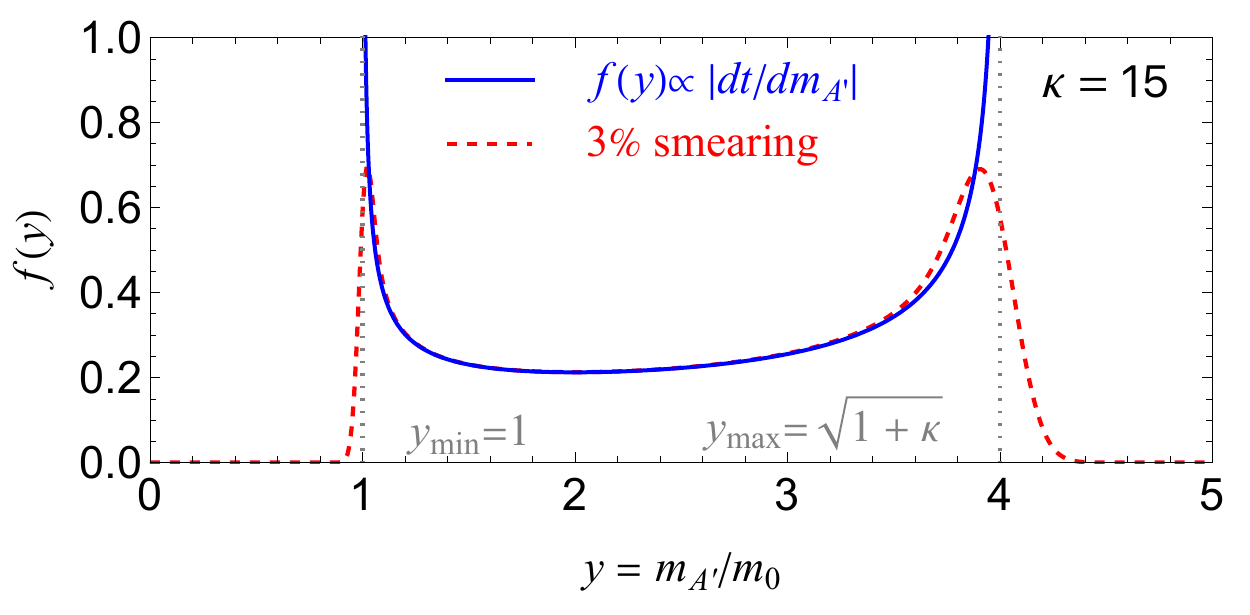}
	\caption{{\bf The normalized probability density function (PDF).} The blue solid and red dashed lines represent the PDF  before and after smearing with a detector resolution of $3\%$, respectively.
	}
	\label{fig:y-pdf}
\end{figure}

The mass ratio $y(t)\equiv m_{A'}(t)/\tilde{m}_0$ has minimum $y_{\min}=1$ and maximum $y_{\max} = \sqrt{1+\kappa}$ respectively, and oscillates with time period of $\tau \equiv \pi/m_\phi$.
From Eq.~(\ref{eq:time-interval}), the invariant mass bin has a time exposure proportional to 
\begin{equation}\label{eq:dtdm}
	\left| \frac{d t}{d m_{A'}} \right|  = \frac{\tau}{ \tilde{m}_0} f(y) , 
\end{equation}
where a factor of 2 is multiplied since each mass appears twice in one period. The probability density function (PDF), 
\begin{equation}\label{eq:f(y)}
	f(y ) =\frac{2 y}{\pi\sqrt{\left(y^2 - y^2_{\min}\right) \left(y^2_{\max} - y^2 \right) }},
\end{equation}
is normalized between $y\in[y_{\min},~y_{\max}]$. Indeed, the time exposure diverges at the minimum and maximum of the resonant mass bin in Fig.~\ref{fig:y-pdf}. After including the detector resolution, it becomes finite and shows the double-peak feature. The right peak contains a larger probability than the left because $f(y)\propto y$. One can evaluate the probability difference contained in the two peaks,
\begin{align}
	\frac{\int_{y_{\max} -\Delta}^{y_{\max} } f(y) d y}{\int_{y_{\min}}^{y_{\min} + \Delta} f(y) d y} \xrightarrow{\Delta \to 0}  \sqrt{\frac{y_{\max}}{y_{\min}}},
	\label{eq:ratio}
\end{align} 
which is a factor of 2 difference for $\kappa \sim \mathcal{O}(15)$.

\begin{table*}
\footnotesize\renewcommand\arraystretch{1.5}\centering
\begin{tabular}{|c|c|c|c|c|c|c|}\hline
Collaboration & Production mode & Experimental environment & Spectrum & Resolution $\sigma_{\rm re}$ & Fit window & Range of $m_{A'}$ \\ \hline
BaBar~\cite{BaBar:2014zli} & $e^+e^-\to\gamma A'$ & $\sqrt{s}\approx10$ GeV, 514 fb$^{-1}$ & $m_{ee}$, $m_{\mu\mu}$ & $[1.5,~8]$ MeV & $m_{A'}\pm10\,\sigma_{\rm re}$ & $[0.02,~10.2]$ GeV \\ \hline
LHCb~\cite{LHCb:2017trq,LHCb:2019vmc,LHCb:2020ysn} & $pp\to A'$ & $\sqrt{s}=13$ TeV, $\sim5~{\rm fb}^{-1}$ & $m_{\mu\mu}$ & $[0.12,~380]$ MeV & $m_{A'}\pm12.5\,\sigma_{\rm re}$ & $[0.214,~69.8]$ GeV \\ \hline
A1~\cite{Merkel:2014avp} & $e^-Z\to e^-ZA'$ & $E_e\in[0.180,0.855]$ GeV & $m_{ee}$ & 0.5 MeV & $m_{A'}\pm3\,\sigma_{\rm re}$ & $[0.040,~0.300]$ GeV \\ \hline
NA48/2~\cite{NA482:2015wmo} & $\pi^0 \to \gamma A^\prime$ & $1.69 \times 10^{7}$ $\pi^0 \to \gamma e^+ e^-$ events & $m_{ee}$& $[0.16,~1.33]$ MeV & single bin & $[0.009,~0.120]$ GeV \\
\hline
\end{tabular}
\caption{The summary table for experiments using the dilepton resonance to search for $A'$. 
}
\label{tab:dilepton}
\end{table*}

If the data taking duration lasts much longer than the oscillation period, $t_{\rm exp}\gg \tau$, the events will run between $\tilde m_0$ and $\sqrt{1+\kappa}\tilde m_0$ many times. In this case, the normalized mass spectrum $f(y)$ fully describes the data distribution without explicit dependence on $t$, initial oscillation phase, or $m_\phi$.  
Since Lyman-alpha constraints require $m_\phi \gtrsim 2 \times10^{-20} ~{\rm eV}$~\cite{Rogers:2020ltq} suggesting the longest oscillation period of about one day, most of the experiments satisfy the $t_{\rm exp}\gg \tau$ condition.

For simple connection between ultraviolet (UV) complete model parameters ($m_0$, $m_\phi$, $g'Q_\phi$) and phenomenological parameters ($m_{A'}$, $\tilde m_0$, $\kappa$), we assume $\arg[\phi_1] = \arg[\phi_2]$ or $\phi_2 =0$ to have 
\begin{align}
	&\tilde{m}_0 = m_0, \\ &\kappa  \equiv 2 (g' Q_\phi)^2 \rho_{\rm DM}/\left(m_\phi^2 m_0^2\right) =  10 \left( \frac{\rho_{\rm DM}}{0.3~\mathrm{GeV}/\mathrm{cm}^3 } \right)   \nonumber \\
	& \times \left(\frac{g' Q_\phi}{1.5 \times 10^{-8}}  \frac{10^{-19}~\mathrm{eV}}{m_\phi} \frac{0.1~\mathrm{GeV}}{m_0}\right)^{2}, \nonumber
\end{align}
which illustrate the possible values of UV parameters in the parameter space of phenomenological interest. Moreover, we are interested in parameter space $m_{0} \sim \mathcal{O}(0.1)$ GeV and $\kappa \sim \mathcal{O}(10)$, connecting to luminosity frontier experiments. 
Normally, the dark photon $A'$ with muon coupling can contribute to $\left(g-2\right)_\mu$ as
$\Delta a_\mu(m_{A'})$. In the time-varying mass scenario, one should average over time as 
\begin{align}
	\frac{1}{\tau} \int_{0}^{\tau} dt ~ \Delta a_\mu\left(m_{A'}(t)\right) .
	\label{deltaal}
\end{align}
Since the varying mass is larger than $m_0$, it generally needs larger $\epsilon$ to explain the $\left(g-2\right)_\mu$ anomaly. 
In addition, the limit from $\left(g-2\right)_e$, $\Delta a_e<0.98\times10^{-12}$ ($95\%$ C.L.) \cite{Morel:2020dww}, need to be revised according to \Eq{deltaal}.

\subsection*{Recasting via the double-peak method (DPM)}

The dark photon has been searched for in the dilepton channels $A'\to\ell^+\ell^-$ ($\ell=e$, $\mu$) with various production mechanisms~\cite{BaBar:2014zli, Merkel:2014avp, NA482:2015wmo, LHCb:2017trq, LHCb:2019vmc, LHCb:2020ysn}. The general strategy is to fit the dilepton invariant mass spectrum $m_{\ell\ell}$ with a given signal hypothesis in a mass window broader than a few times of the energy resolution $\sigma_{\rm re}$. The upper limits on the production cross section can be translated into bounds on $\epsilon^2$ as a function of $m_{A'}$.  To recast the BaBar and LHCb results, for a given mass window, we fit the $m_{\ell\ell}$ spectrum with a quadratic or cubic function and compare it to the observed data with and without the signal events to get the likelihoods $\mathcal{L}$ and $\mathcal{L}_0$, respectively. Then we require the log-likelihood ratio $\rm{LLR}\equiv -2\ln(\mathcal{L/L}_0) = 3.84$ to obtain the upper limit for signal event number corresponding to 95\% confidence level at rejecting a signal hypothesis~\cite{Cowan:2010js}.

\begin{figure}
	\centering 
	\includegraphics[width=0.48 \textwidth]{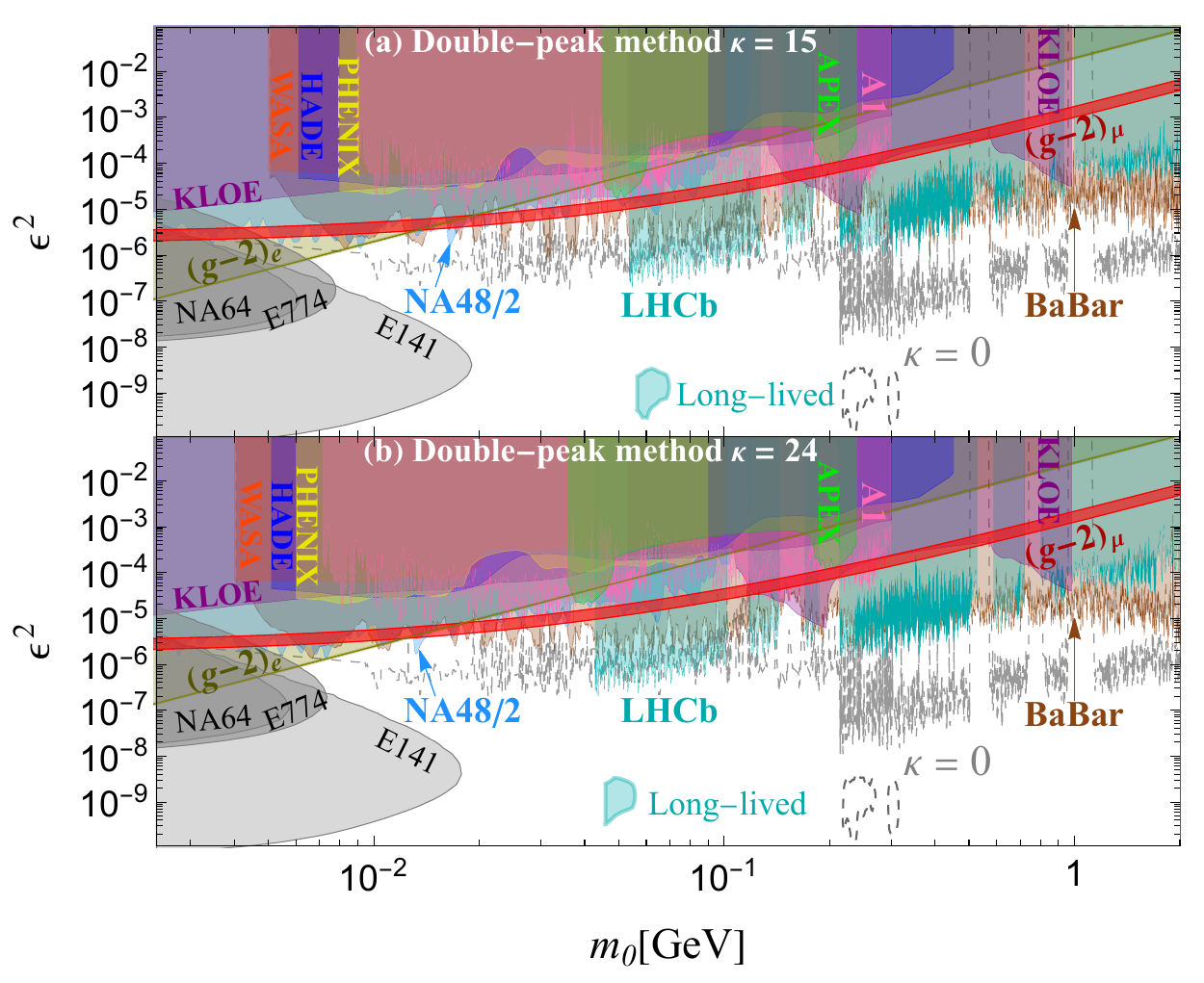}
	\caption{{\bf Limits on mixing strength $\epsilon^2$ as a function of mass parameter $m_0$.} The colored shaded regions in (a) and (b) represent the excluded parameter space using the double-peak method for $\kappa = 15$ and $24$, respectively, and the gray dashed line represents the traditional limit ($\kappa =0$). The red-shaded region can provide a solution to $\left(g-2\right)_\mu$.
	}
	\label{fig:kappa}
\end{figure}

We first recast the traditional single-peak resonance signal method for dilepton experiments following the experimental setups summarized in Table~\ref{tab:dilepton}. Our recast results agree with the experimental results quite well, and the detailed process is in Methods and Supplemental Notes 2 and 4 \cite{BaBar:2014zli, LHCb:2020ysn, LHCb:2019vmc, Merkel:2014avp, Bross:1989mp, Riordan:1987aw, NA64:2019auh, Andreas:2012mt, Bjorken:2009mm, Bauer:2018onh, Liu:2020qgx}.
We next apply our time-varying resonance signal model to fit the background data. According to the double-peak feature of the signal, for a fixed mass window centering around $m_{A'}$, there are two signal peaks to fit, the minimum $m_{A'}=m_0$ and the maximum $m_{A'}=\sqrt{1+\kappa}m_0$. Thus, one can obtain two sets of $\epsilon^2$ constraints as a function of $m_0$, and the stronger one is adopted as the $\epsilon^2$ limit for a given $m_0$. For a given $m_0$, the best limit usually comes from the maximum. Note that one can even constrain $m_0$ below the dilepton mass threshold via the maximum peak; e.g., the dimuon limit of LHCb extends to $m_0$ much smaller than $2M_\mu$, as shown in Fig. \ref{fig:kappa}.

\begin{figure}
	\centering 
	\includegraphics[width=0.48 \textwidth]{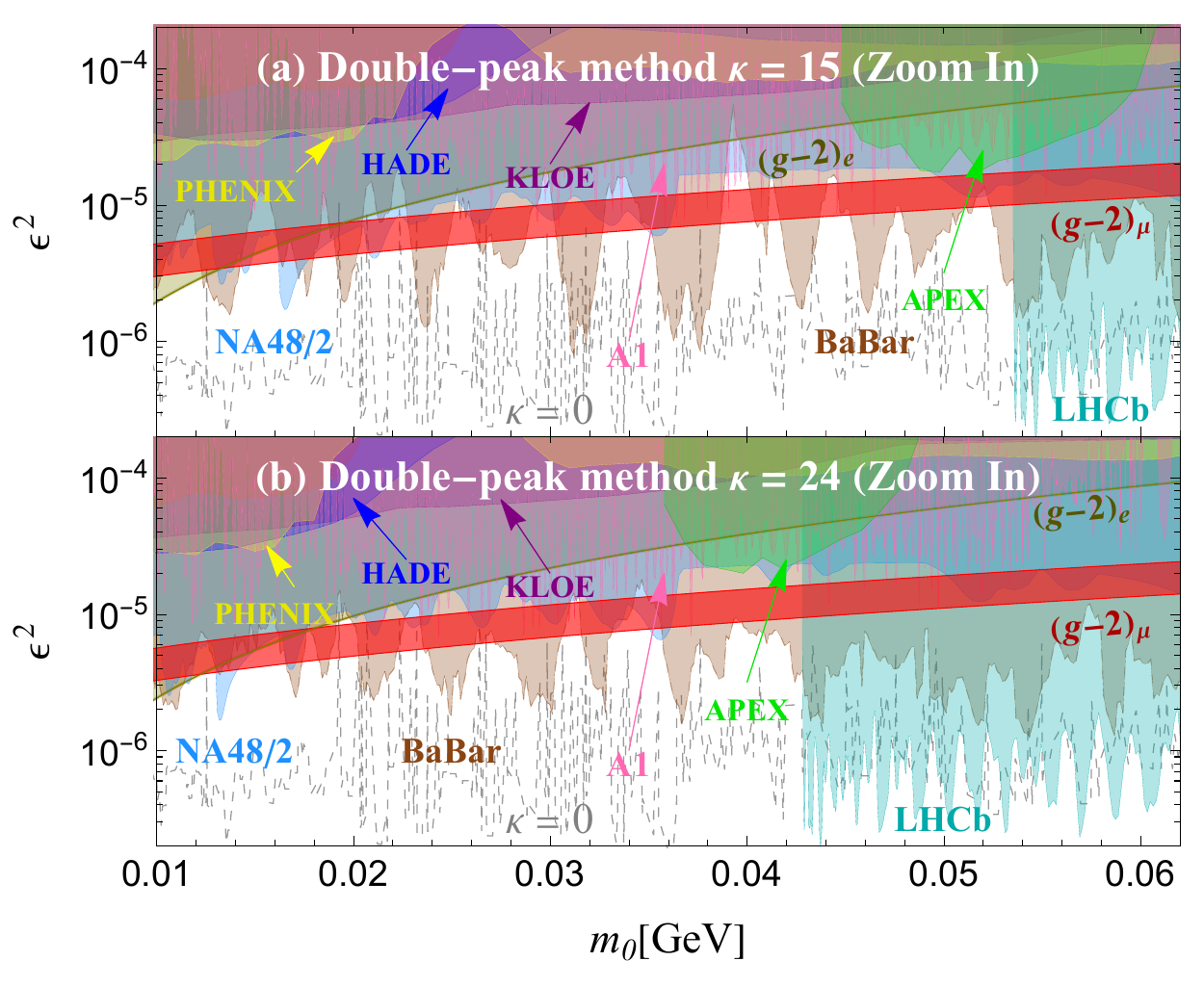}
	\caption{{\bf Limits on mixing strength $\epsilon^2$ in the mass range $m_0\in[0.01, 0.07]$~GeV.} The colored shaded regions in (a) and (b) represent the excluded parameter space using the double-peak method for $\kappa = 15$ and $24$, respectively, and the gray dashed line represents the traditional limit ($\kappa =0$). The red-shaded region can provide a solution to $\left(g-2\right)_\mu$.
	}
	\label{fig:zoomin}
\end{figure}
In Fig.~\ref{fig:kappa}, we apply LLR calculation to experiments in Table~\ref{tab:dilepton}. Other experiments such as APEX~\cite{APEX:2011dww}, HADES~\cite{HADES:2013nab}, KLOE~\cite{KLOE-2:2016ydq,Anastasi:2015qla,KLOE-2:2014qxg,KLOE-2:2012lii}, PHENIX~\cite{PHENIX:2014duq} and WASA~\cite{WASA-at-COSY:2013zom}, generally provide relatively weaker constraints on $\epsilon^2$ in interested parameter space. Hence, we recast their results by rescaling the limits on $\epsilon^2$ according to the time exposure in a bin. We checked the robustness of such a simplified method by applying it to the LHCb experiment and obtaining a good agreement with the full-fitting LLR method, whose details are discussed in Supplemental Note 1.

The constraints on our time-varying signal model from the above experiments are plotted in Fig.~\ref{fig:kappa}. For $m_0\gtrsim10^{-2}$ GeV and $\kappa\sim\mathcal{O}(10)$, current experiments constrain $\epsilon^2\gtrsim10^{-7}-10^{-5}$, around one order weaker than the traditional single-peak bounds, whose envelope is shown as gray dashed line labeled as $\kappa=0$. Especially, the excluded $(g-2)_\mu$ solution becomes viable at $\mathcal{O}(10)$ MeV when the $(g-2)_\mu$ red band crosses with the BaBar NA48/2 bounds, as depicted in Fig. \ref{fig:zoomin}.

There are beam dump experiments E774~\cite{Bross:1989mp}, E141 \cite{Riordan:1987aw} and NA64 \cite{NA64:2019auh}. They set limits on $A' \to\ell^+\ell^-$ based on the signal event number $N(\epsilon, m_{A'})$, for given $\epsilon$ and $m_{A'}$.
The $A' $ is produced at beam dump and propagates a distance according to its lifetime and decay to $\ell^+ \ell^-$.  We can translate the  upper limit on the event 
number of $A'$ decay, e.g., 17 events for E774, to our scenario by simply time averaging the signal events as
\begin{align}
	\frac{1}{\tau}\int_0^\tau N(\epsilon, m_{A'}(t))dt ,
\end{align}
then compare it with the upper limit as shown in Fig.~\ref{fig:kappa}. The detailed estimation of $N(\epsilon, m_{A'})$ is given in Methods. In addition, the LHCb limits for the displaced $A' $ are shown in Fig.~\ref{fig:kappa} with details given in Supplemental Note 2.
For $A'$ invisible decaying \cite{BaBar:2017tiz, Zhang:2019wnz, Banerjee:2019pds, NA62:2019meo}, the double-peak method can apply in the same way.

\begin{figure}[htbp]
	\centering 
	\includegraphics[width=0.45 \textwidth]{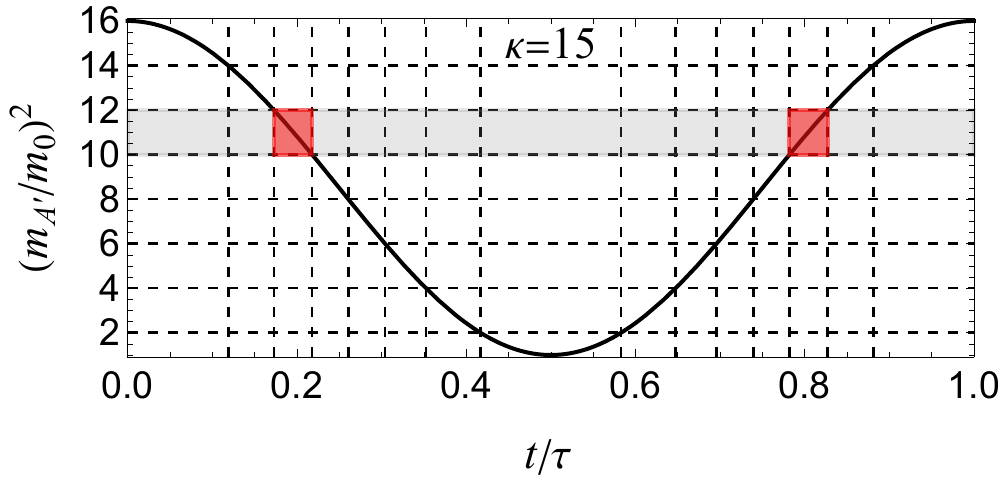}
	\caption{{\bf Time grids used in the time-dependent method.} The resonant mass curve is represented by the black solid line. For a fixed row (gray), two particular bins (red) are chosen in the time-dependent method.}
	\label{fig:TDM}
\end{figure}

\subsection*{Improving by the time-dependent method (TDM)}
 
Previous calculations only consider two parameters, $\kappa$ and $m_0$, while do not exploit the recorded time-stamp of events, $t$.
The experiments can reanalyze the data using all the above three parameters,
because the signal events only happen at certain time $t$ and mass $m_{A'}(t)$,
as shown in Fig.~\ref{fig:TDM}. In principle, the experiments can figure out $\kappa$, $m_0$, $m_{\phi}$ and the initial phase via a two-dimension fit on the $t$-$m_{\ell\ell}$ plane; for example, dark matter mass $m_{\phi}$ can be extracted by the the period of mass modulation of dark photon via $\tau=\pi/m_\phi$, while the dark photon bare mass $m_0$ and $\kappa$ can be extracted via the minimum and maximum of signal $m_{\ell\ell}$ distribution, etc. Note that a direct probe of the ultralight dark matter mass $m_\phi$ is possible only in TDM through the time-varying feature of the dark mediator $A'$.

\begin{figure}[htbp]
	\centering 
	\includegraphics[width=0.48 \textwidth]{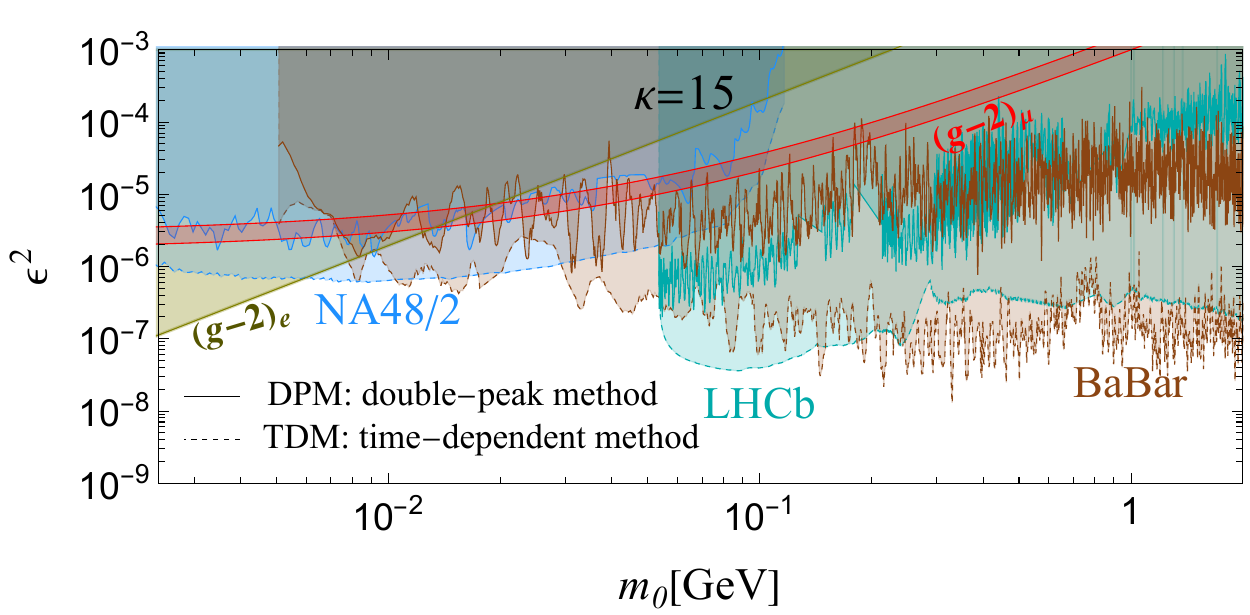}
	\caption{{\bf The comparison between double-peak and time-dependent methods.} The constraints and projected reach for BaBar \cite{BaBar:2014zli}, LHCb~\cite{LHCb:2017trq,LHCb:2019vmc,LHCb:2020ysn} and NA48/2 \cite{NA482:2015wmo} are shown in colored shaded regions, and the $\left(g-2\right)_\mu$ solution region is covered by red.
	}
	\label{fig:lhcb15}
\end{figure}

Without the time information, we can assume the observed data has a flat probability in time and estimate the signal sensitivity.
We adopt the same mass grid as the experiment, and it automatically generates the time grid according to the signal mass function $m_{A'}(t)$.
Specifically, if the total number in the $i$th mass bin is $N_i$, we have the number of data in $i$th mass bin and $j$th time bin as $N_{i, j} = N_{i} \Delta t_j/\tau$.
For a fixed mass bin (horizontal gray shaded), we pick up the two red bins in Fig.~\ref{fig:TDM},  which contain the signal. Adding the data in red bins together, we have
\begin{align}
	N_{i}^{\rm red} 
	= N_i \frac{1}{\tau} \int_{m_i}^{m_{i+1}}  \left | \frac{dt}{dm_{A'}}\right | dm_{A'},
\end{align}
forming an updated set of data $N_i^{\rm red}$. Then, our previous calculation using the DPM can apply. 

In this method, the signal event number is unaffected while the background event is suppressed by a factor of $\frac{1}{\tau} \int_{m_i}^{m_{i+1}}  \left | \frac{dt}{dm_{A'}}\right | dm_{A'}$.
Fig.~\ref{fig:lhcb15} shows the projected sensitivities using TDM for NA48/2, BaBar, and LHCb as examples, assuming no excess is detected. Indeed it improves the reach by $1$--$2$ orders of magnitude compared with the DPM, and the viable regions of $(g-2)_\mu$ at $\mathcal{O}(10)$ MeV in Fig.~\ref{fig:kappa} can be probed now. We then propose the experimentalists to reanalyze the data using TDM to probe the time-varying signal model. Reducing the invariant mass resolution and hence the size of $m_{\ell\ell}$ bin can reduce the number of background events in the corresponding mass bin, which enhances the sensitivity to signal, as the number of signal events is unaffected when picking up the red bins along the oscillation trajectory in Fig.~\ref{fig:TDM}. However, if the mass bin is too small, the analysis will suffer from statistical error due to small $N_i^{\rm red}$.

\subsection*{The invisible dark photon}

Due to small $g' Q_{\phi}$, the dominant decay channels of $A'$ are SM fermions in our minimal setup. However, in an extended dark sector, $A'$ may decay to invisible particles dominantly. For example, if some dark fermion $\chi$ that carries dark $U(1)'$ charge, then $A'\to\chi\bar\chi$ channel can dominate the $A'$ decay with the subsequent decay of $\chi$ to DM. In that case we have an invisible decaying dark photon scenario, and it has been searched by several experiments including BaBar \cite{BaBar:2017tiz}, BES-III \cite{Zhang:2019wnz}, NA64 \cite{Banerjee:2019pds}, and NA62 \cite{NA62:2019meo}. We will briefly discuss how the time-varying scenario can affect the results. BaBar and BES-III have studied the monophoton channel, $e^+ e^- \to A^\prime \gamma$ with invisible $A^\prime$. The photon energy is $E_\gamma = (s-m_{A^\prime}^2)/(2 \sqrt{s})$ with $\sqrt{s}$ being total collision energy. With time-varying $m_{A'}(t)$, $E_\gamma $ extends to a spectrum determined by the following differential, 
\begin{align}
	\left| \frac{d t}{d E_\gamma} \right|  =  \frac{\tau}{\pi\sqrt{\left(E_\gamma - E_{\min}\right) \left(E_{\max} -E_\gamma \right) }} ,
	\label{eq:dtdE}
\end{align}
where $E_{\min} \equiv (s-(1+\kappa)m_{0}^2)/(2 \sqrt{s})$ and $E_{\max} \equiv  (s-m_{0}^2)/(2 \sqrt{s})$. Then, the analysis is similar to visible $A'$ by substituting the invariant mass bin for the photon energy bin, and it is expected to weaken the limit. The exception happens for very small $m_0$ satisfying $m_0^2 < 2 \sqrt{s} \sigma_\gamma/\kappa$, with $\sigma_\gamma$ being the photon energy resolution. In this case, the limit will be unchanged because both $E_{\min}$ and $E_{ \max}$ fall into the same photon energy bin.

\begin{figure}
	\centering 
	\includegraphics[width=0.48 \textwidth]{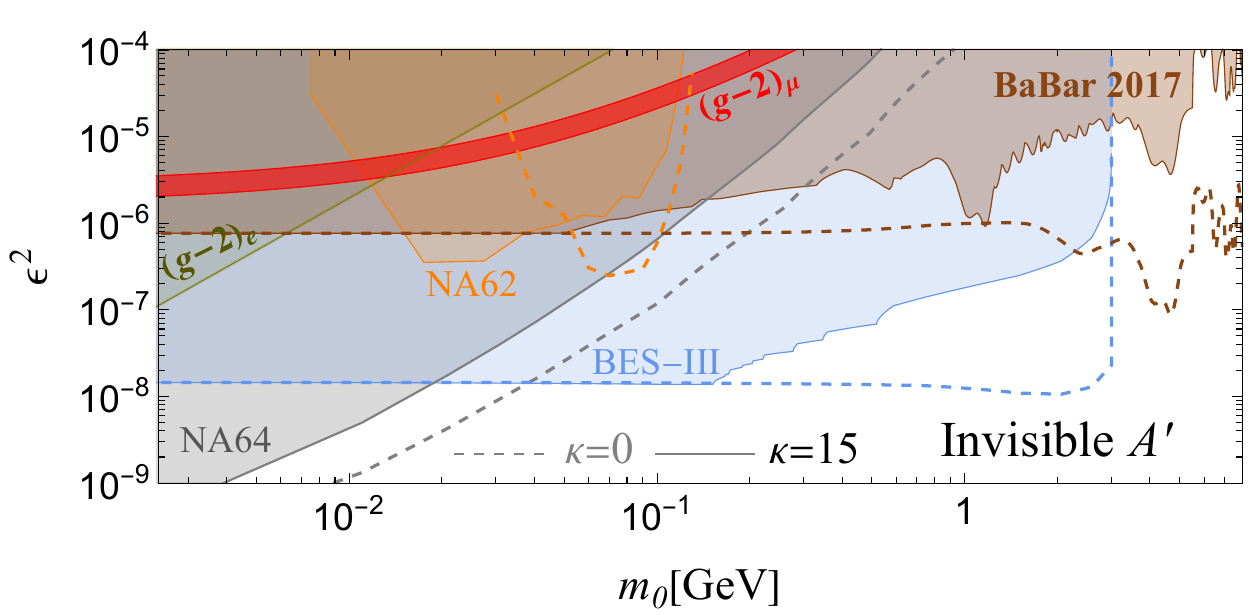}
	\caption{{\bf Limits on mixing strength $\epsilon^2$ from invisible dark photon searches.} The colored regions are constraints derived by double-peak method for $\kappa = 15$ for invisible dark photon searches (BaBar \cite{BaBar:2017tiz}, BES-III \cite{Zhang:2019wnz}, NA64 \cite{Banerjee:2019pds} and NA62 \cite{NA62:2019meo}).
	}
	\label{fig:invisible11}
\end{figure}

The electron beam dump experiment NA64 \cite{Banerjee:2019pds} studies the process $e^- Z \to e^- Z A^\prime$ for invisible $A^\prime$. The signal events are selected with $E_{\rm miss} > 50$ GeV and other cuts on electromagnetic and hadronic calorimeter energies. Since $E_{\rm miss} $ is much larger than interested $m_{A'}$, the cut efficiency should not significantly depend on $m_{A'}$.
Therefore, the dominant effect of time-varying $m_{A'}$ will show up in the production rate, scaling as $m_e^2/m_{A'}^2$ \cite{Bjorken:2009mm}. Thus one can take the time average for this factor to estimate the weakening of the limits.
Another beam dump experiment NA62~\cite{NA62:2019meo} focuses on invisible $A'$ from $\pi^0 \to \gamma A'$, where $A'$ mass is reconstructed using the process $K^{\pm} \to \pi^\pm \pi^0 $ as $m_{\rm res}^2 = (p_{K^\pm} - p_{\pi^\pm}-p_\gamma)^2$. The double-peak method can analyze the data and set updated limits. 

The results are shown in Fig. \ref{fig:invisible11}, in which the $\left(g-2\right)_\mu$ parameter space has been entirely excluded, even for the time-varying signal. Therefore, we shall assume the invisible decay channel is subdominant to have a dark photon $\left(g-2\right)_\mu$ solution.

\begin{figure}[htbp]
	\centering 
	\includegraphics[width=0.48 \textwidth]{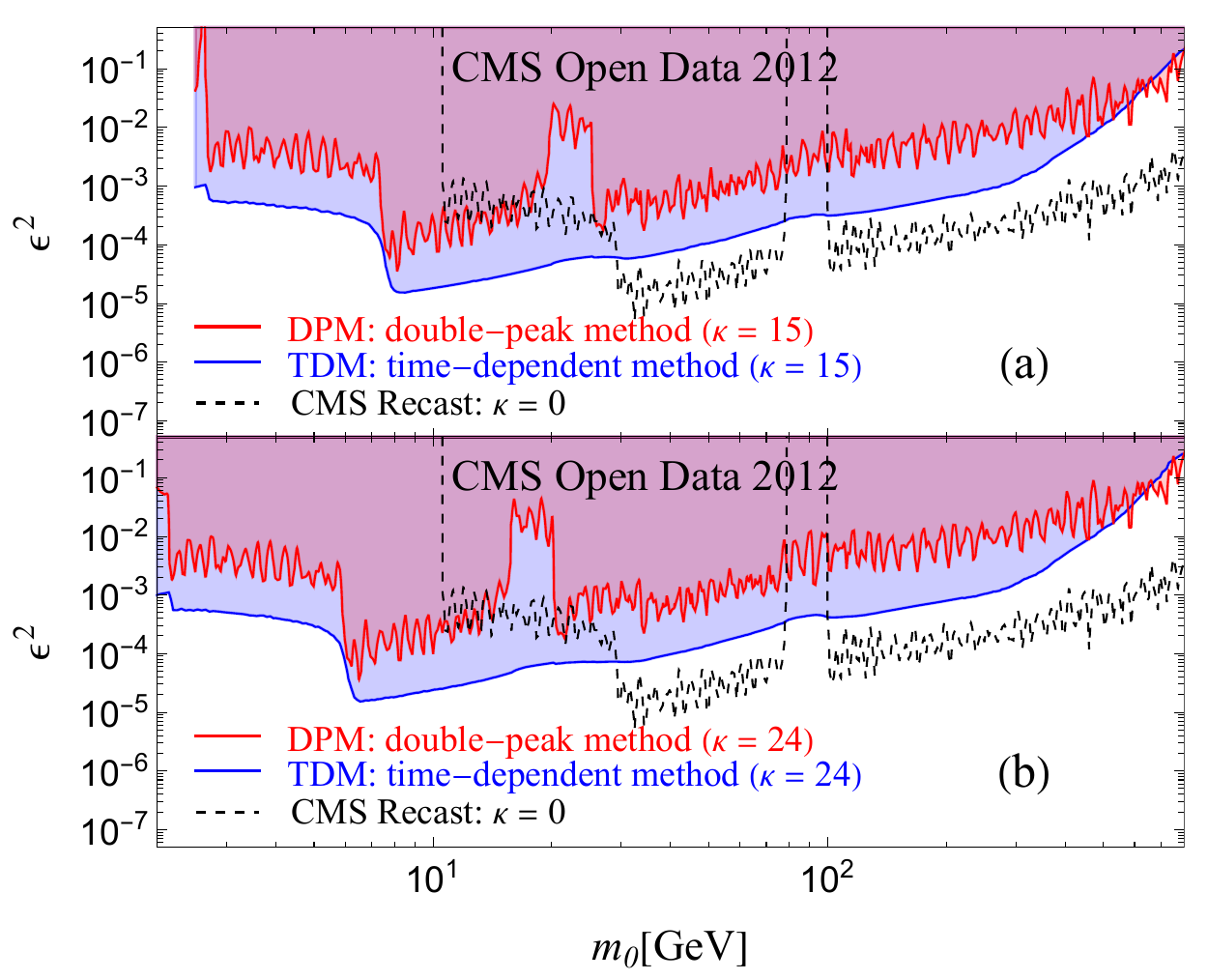}
	\caption{{\bf Limits on mixing strength $\epsilon^2$ from CMS Open Data.} The initial phase is $\phi_0=0$ and the DM mass is $m_\phi = 10^{-19}$ eV, while $\kappa=15$ (a) and 24 (b). We show the recast of traditional limit ($\kappa =0$) in gray dashed, and DPM and TDM limits in purple and blue.}
	\label{fig:CMS}
\end{figure}

\subsection*{Analyzing CMS Open Data}
 
The 2012 dimuon events from CMS Open Data~\cite{CMSDoubleMu2012RunB, CMSDoubleMu2012RunC} can be used to justify the two analyzing methods, DPM and TDM. We recast the CMS analysis with the traditional single-peak model shown as $\kappa = 0$, in Fig.~\ref{fig:CMS}.

For the real data, there is a non-trivial feature: the instant luminosity $\mL(t)$ is non-uniform. It introduces time dependence from human operation aside from the time oscillation from the signal. Therefore, the invariant mass distribution becomes
\be 
\frac{dN_S}{dm_{\ell\ell}}= \frac{\epsilon_S\sigma_0 }{m_0}  \times f\left(\frac{m_{\ell \ell}}{m_0}\right)
\frac{\tau}{2 }\sum_{{i}} \left[ \mL(t_{i}^+) + \mL(t_{i}^-) \right],
\label{eq:instantL}
\ee
where $\epsilon_S$ and $\sigma_0$ are the cut efficiency and the production cross-section at $m_{A'} = m_{\ell \ell}$, and $t_i^\pm$ are the two solutions in the $i$-th time period satisfying $m_{A'}(t) = m_{\ell \ell}$. When the oscillation period $\tau$ is smaller than the typical $\mL(t)$ variation time (about a few hours), $\mL(t)$ becomes a constant in each period. Therefore, $\frac{\tau}{2 }\sum_{{i}} \left [ \mL(t_{i}^+) + \mL(t_{i}^-) \right] = \text{L} $ and we can apply DPM to CMS Open Data. 
In addition, we can also apply TDM to CMS Open Data after filling the data into $t$-$m_{\ell\ell}$ grids. More details for the analysis of CMS Open Data are given in Methods \cite{CMS:2011xlr, cms2012cms, Curtin:2014cca, CMS:2014lcz, Lassila-Perini:2021xzn}.

The results from DPM and TDM are plotted in Fig.~\ref{fig:CMS} for $\kappa=15,~24$, $\phi_0=0$, and $m_\phi = 10^{-19}$ eV. As expected, the limits from DPM are weaker than our recast for the traditional limit ($\kappa = 0$). Moreover, the limits for TDM are indeed better than DPM by $1$--$2$ orders. We show that TDM can work for the real collider data.

\subsection*{Other constraints}

Besides collider and beam dump experiments, there are other constraints to clarify. 
Firstly, the coupling $g' Q_\phi $ should be small to avoid $\phi$ thermalization via self-scattering and scattering with normal matter via $f \phi \to f \phi$ or $f \bar{f} \leftrightarrow \phi \phi^*$.
Secondly, the $A' \to \phi \phi^*$ decay could freeze-in $\phi$ as a hot relic, which should be very small.
For interested parameter space $m_0 \sim \mathcal{O}(0.1)$ GeV, the coupling $g' Q_{\phi}$ around $10^{-6} \text{--} 10^{-10}$, is small enough to satisfy the above requirements for $m_\phi \sim 10^{-20}\text{--}10^{-17}$ eV, while keeping $\kappa \sim \mathcal{O}(10)$. 
Moreover, in the early Universe, the field value of $\phi$ is much larger than today. Thus, a heavy $A'$ mass helps to evade the thermalization and freeze-in constraints.
For the ultralight scalar, the black hole super-radiance can exclude some mass regions, but not all the interested regions \cite{Arvanitaki:2014wva, Davoudiasl:2019nlo}.

At 1-loop level, the SM fermion mass can receive a QED-like correction from $A'$ interaction,
\begin{align*}
	\frac{\Delta m_f}{m_f}\simeq \frac{3 \left(e \epsilon Q_f \right)^2 }{16\pi^2} \log\left(\frac{m_{0}^2+ 2(g'Q_\phi)^2 \phi^*\phi}{m_0^2}\right) ,
\end{align*}
with $m_{A'}\gg m_f$. This leads to a logarithmic coupling between $\left(\phi^* \phi\right)$ and the fermion mass operator. Its Taylor expansion can not be naively truncated due to large $\kappa$, hence is different from the linear and quadratic couplings. Especially for experiments relying on certain $\phi$ field distribution around massive objects, such as Cassini stochastic, binary pulsars tests \cite{Armstrong:2003ay, Blas:2016ddr}, atomic clocks \cite{VanTilburg:2015oza, Hees:2016gop}, torsion balances \cite{Smith:1999cr, Schlamminger:2007ht, Wagner:2012ui}, and MICROSCOPE space experiment \cite{Touboul:2017grn}, the constraints do not simply apply. As for the fifth force experiments \cite{Adelberger:2006dh, Kapner:2006si}, even for the quadratic coupling, it only provides a loose bound \cite{Olive:2007aj, Stadnik:2015kia}, which can be easily satisfied for small $\epsilon $ and $g'Q_\phi$. The constraints from Big Bang nucleonsynthesis due to the enhanced $\phi$ value \cite{Stadnik:2015kia, Sibiryakov:2020eir, Chen:2019kcu}, can be easily evaded in our scenario thanks to the logarithmic coupling.

\section*{Conclusion}

We have introduced an innovative strategy to search for ultralight bosonic DM, namely the time-varying resonance from the dark sector at the collider and beam dump experiments. We found it can lead to double-peak feature in the invariant mass spectrum, which can help to evade the dilepton and missing mass resonant searches at collider and beam dump experiments. Moreover, the mass spectrum is independent of time as long as the oscillation period is short compared with the variation time scale of instant luminosity. A concrete model is discussed with ultralight complex scalar DM inducing an oscillating mass for kinetic mixing dark photon. For mass around tens of MeV, the already excluded muon $\left(g-2\right)_\mu$ solution from $A'$ becomes viable again; this parameter region can be further tested by reanalyzing the existing data with event-by-event time stamps, as shown by the time-dependent method.
We use CMS Open Data for a time-dependent resonance search and justify that our method works as expected, even with the complexity of a non-uniform instant luminosity. We also demonstrate its application in the case of invisible decay dark mediator at the electron-positron collider.
In general, the analysis strategy can search the time-varying signal from the ultralight DM oscillation at collider and beam dump experiments.

%
\section*{Methods}




We show the detailed log-likelihood ratio (LLR) calculations for the collider and beam dump experiments. Next, the recasts for the existing experiments are in good agreement with the official results. Therefore our calculations using double-peak (DPM) and time-dependent methods (TDM) are robust.
Aside from providing projected limits for existing experiments, we use CMS Open Data to provide real collider limits using DPM and TDM. Lastly, we constrain time-varying $A'$, which decays to the invisible final state using the DPM.

\subsection*{The LLR calculation for collider experiments}

In this section, we will show the details of LLR calculation which sets limits on dark photon signal cross section or mixing parameter $\epsilon$. Taking the BaBar experiment as an example, we recast the limits of the traditional single-peak resonance model ($\kappa = 0$) and compare them with official results. Good agreements are obtained. We then adopt the time-varying signal model and derive the updated bounds for both the double-peak and time-dependent analysis strategies. The details of the recast and updated bounds of LHCb, and A1 experiments are given in Supplemental Notes 2, 3 and 4, including the long-lived dark photon at the LHCb experiment for the time-varying scenario.

The BaBar collaboration collected 514 fb$^{-1}$ data at the vicinity of the $\Upsilon(4S)$, $\Upsilon(3S)$ and $\Upsilon(2S)$ resonances to search for $e^+e^-\to\gamma A'$ process, with $A'\to e^+e^-$ and $\mu^+\mu^-$ decay channels within a mass range $0.02~{\rm GeV}<m_{A'}<10.2~{\rm GeV}$~\cite{BaBar:2014zli}. A total of $N_e=5704$ ($N_\mu=5370$) mass hypotheses are searched in the $e^+e^-$ ($\mu^+\mu^-$) channel respectively. For a given mass $m_{A'}$, an interval of $m_{A'} \pm 10\,\sigma_{\rm re}$ is used to perform the fits, where $\sigma_{\rm re}$ is the energy resolution at $m_{A'}$, varying from 1.5 to 8 MeV in the whole $m_{A'}$ range. Even though the full data is not given in Ref.~\cite{BaBar:2014zli}, there are available data of  $m_{ee}$ and $m_R=\sqrt{m_{\mu\mu}^2-4M_\mu^2}$ spectrum up to $\sim10$ GeV with a uniform bin size of 100 MeV in the main text, and the zoomed-in spectrum for $m_{ee}\in[17.8,62.2]$ MeV ($m_R\in[0.522,77.4]$ MeV) with a bin size of 0.5 MeV (1.0 MeV) in the appendix. We refer to the former and latter as the low- and high-granularity datasets, respectively.

We first generate the artificial high-granularity data for the whole mass spectrum to recast the BaBar results. For $m_{ee}<62.2$ MeV ($m_R<77.4$ MeV), we adopt the high-granularity data themselves, while for higher mass, we use the interpolation of the low-granularity data, including appropriate statistical smearing. We assume the bin size increases linearly with resonant mass and keep the total number of the data to be $N_e$ ($N_\mu$), respectively. Moreover, we assume the resolution $\sigma_{\rm re}$ increases linearly with the resonant mass. Finally, for a given mass point $m_{A'}$, we fit the artificial data in the mass window $m_{A' } \pm 10\, \sigma_{\rm re}$. 
The LLR is defined as
\begin{equation}\label{eq:LLR}
	{\rm LLR}=-2\log\left[\frac{{\rm Max}_{\vec{a}'}\prod_i\mathcal{N}(B_i-B(m_i,\vec{a}')-S f_G(m_i)|B_i)}{{\rm Max}_{\vec a}\prod_i\mathcal{N}(B_i-B(m_i,\vec{a})|B_i)}\right],
\end{equation}
where $B_i$ is the background event number in the $i$th mass bin, and
\begin{equation}
	B(m_i,\vec{a})=a_0 + a_1 m_i + a_2 m_i^2,
\end{equation}
is the background fitting function with $m_i$ being the center value of $i$th bin for $m_{ee}$ or $m_R$, while
\begin{equation}
	\mathcal{N}(x|\sigma^2)\equiv\frac{1}{\sqrt{2\pi}\sigma}\exp\left\{-\frac{x^2}{2\sigma^2}\right\},
\end{equation}
is the normalized Gaussian distribution and $S$ is the total signal event number. $f_G(m_i)$ is the signal template without the time-varying effect, and after detector smearing, it is defined as
\begin{equation}
	f_G(m_i)=\mathcal{N}(m_{A'} - m_i|\sigma_{\rm re}^2).
\end{equation}
In the LLR calculation, only statistical error is considered, and there is an extra Jacobi factor for the dimuon channel from the definition of $m_R$.

\begin{figure*}[htbp]
	\centering 
	\includegraphics[width= \textwidth]{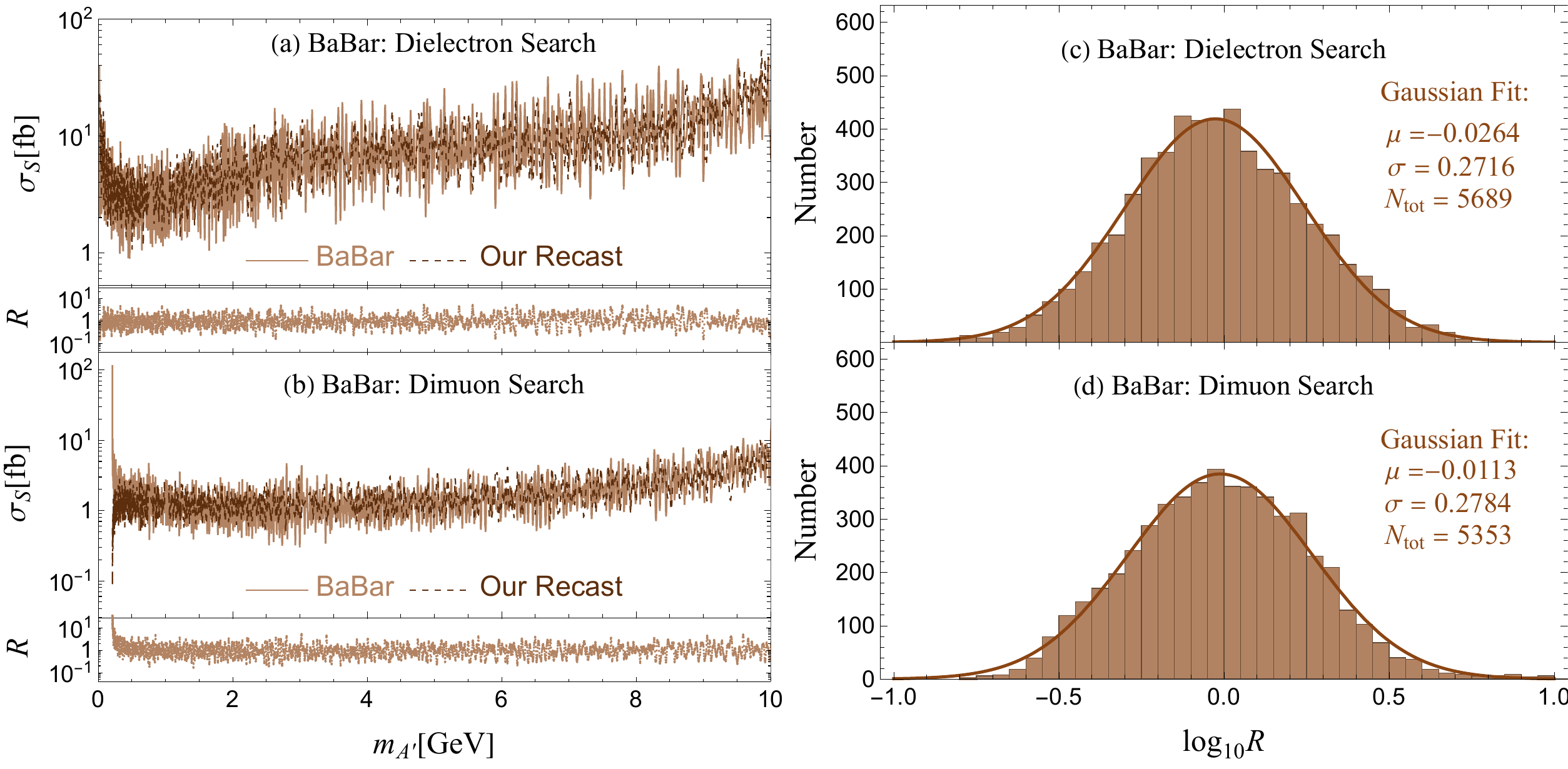}
	\caption{{\bf Recast of the BaBar data.} (a): The upper limit on signal cross-section without the time-varying effect, as a function of $m_{A'}$ in dielectron ($e^+e^-$) and dimuon ($\mu^+\mu^-$) channels for the BaBar experiment~\cite{BaBar:2014zli}. The existing and our recast limits are plotted as solid and dashed. (b): the function $R$ shows the ratio between our recast and official results.
	(c)(d): The distribution of $\log_{10}R$ for $e^+e^-$ and $\mu^+\mu^-$ channels. }
	\label{fig:babar-recast}
\end{figure*}

After requiring ${\rm LLR}=3.84$ (the 95\% confidence level at rejecting a signal hypothesis), we obtain the limits on the allowed signal total event $S$. They can be used to unfold the limit on $\sigma(e^+e^-\to\gamma A')$ via the acceptance factor 0.15 (0.35) in the dielectron (dimuon) channel respectively~\cite{BaBar:2014zli}. 
We found our recast results are consistent with BaBar's official results, as shown in Fig.~\ref{fig:babar-recast}(a) for both $e^+ e^-$ and $\mu^+\mu^-$ channels. To test our recast result, we define a ratio $R= \sigma_S^{\rm Recast}/\sigma_S^{\rm Official}$, which is shown as the low panels of Fig.~\ref{fig:babar-recast}(a). 
In Fig.~\ref{fig:babar-recast}(b), we show the distribution of $\log_{10}R$ for our simulated points. As expected, the $\log_{10}R$ is distributed around 0.  We fit the distribution with a standard Gaussian function $g(x)=\frac{N_{\rm tot}}{\sigma \sqrt{2 \pi}} \exp \left(-\frac{1}{2} \frac{(x-\mu)^2}{\sigma^2}\right)$, where $N_{\rm tot}$ is the total number of points, $\mu$ and $\sigma$ is the mean value and standard deviation respectively. For $e^+e^-$ and $\mu^+\mu^-$ channels, we found similar Gaussian shapes with $\sigma = 0.28$ and $\mu \approx 0$. Therefore, our recasts are quite close to BaBar's official results. Our deviation from the official result is within a factor of $10^{0.28} \approx 1.9$ at $1\sigma$ level.
As a result, our recasts are quite robust. Therefore, our LLR calculation and the projected limits can be trusted.

Regarding the time-varying scenario, the signal invariant mass spectrum follows the probability density function as defined in Eq. (12) of the main text,
\begin{align*}
	f(y)= \frac{2y}{\pi \sqrt{(y^2 - 1)(1+\kappa -y^2)}},
\end{align*}
with $y = m_{\ell\ell}/m_0$ and $y$ takes value from $\left[1, \sqrt{1+\kappa}\right]$. With Gaussian smearing, the signal template becomes
\begin{equation}
	f_S(m_i)=\int_{m_{\min}}^{m_{\max}}f\left(\frac{m'}{m_0}\right)\mathcal{N}(m_i-m'|\sigma_{\rm re}^2)dm',
\end{equation}
with $m_{\min}=m_0$ and $m_{\max} = \sqrt{1+\kappa}m_0$. In the $\mu^+\mu^-$ channel, the additional Jacobi factor should be considered as $m_i$ refers to $m_R$, not $m_{\mu\mu}$. The time-varying scenario is then fitted using $f_S$ in Eq.~(\ref{eq:LLR}). As $f_S$ peaks around $m_{\min}$ and $m_{\max}$, for a given $m_{A'}$ we perform two independent fits for $m_0 = m_{A'}$ and $m_0 = m_{A'}/\sqrt{1+\kappa}$ respectively. Therefore, for a given $m_{A'}$, we can obtain two sets of limits on the allowed $S$ corresponding to the left and right peaks, respectively.

To reduce the systematic uncertainties, we take the ratio of $S$ from the time-varying resonant mass scenario $f_S$ and the traditional resonant scenario $f_G$ to rescale the $\epsilon^2$ constraints from the BaBar measurement~\cite{BaBar:2014zli}. Since the right peak of $f_S$ is usually higher than the left peak, for a given $m_0$, the most stringent constraint usually comes from the mass window around $m_{A'}=\sqrt{1+\kappa}m_0$. Therefore, this is adopted as the BaBar constraints on our time-varying resonance scenario.

\subsection*{The detailed limits calculation for beam dump experiments}

\begin{figure}[htbp]
	\centering 
	\includegraphics[width=0.48 \textwidth]{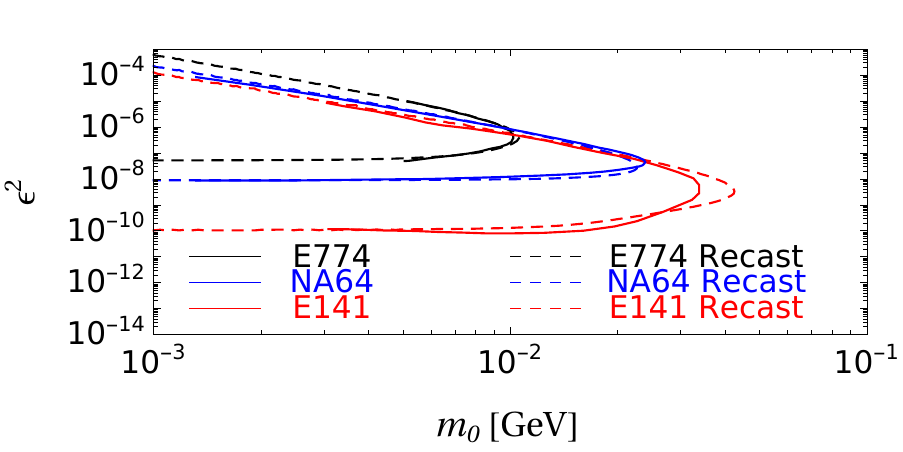}
	\caption{{\bf Recast of the beam dump experiments.} The constraints obtained from our estimation Eq.~(\ref{eq:nob}) are shown in dashed lines, comparing with corresponding results obtained from experiments and early analysis shown in solid lines.}
	\label{fig:deamdumre}
\end{figure}

The beam dump experiments E774~\cite{Bross:1989mp}, E141~\cite{Riordan:1987aw} and NA64~\cite{NA64:2019auh}, are all electron fixed-target experiments. The dark photon $A^\prime$ are dominantly produced by electron Bremsstrahlung process, $e^- Z \to e^- Z A'$~\cite{Andreas:2012mt,Bjorken:2009mm}. The $A^\prime$ will travel some distance and decay before reaching the detector. Since there is shielding behind the collision target, $A^\prime$ must have a lifetime with macroscopic scale. To simplify our analysis, we employ the estimate of the $A^\prime$ signal event number following Refs. \cite{Andreas:2012mt,Bjorken:2009mm,Bauer:2018onh,Liu:2020qgx},
\begin{equation}
	N(\epsilon,m_{A^\prime})= N_e \mathcal C^\prime \epsilon^2 \frac{m_e^2}{m_{A^\prime}^2} e^{-a_1 L_{\text{sh}}\Gamma_{A^\prime}}(1-e^{-a_2 L_{\text{dec}}\Gamma_{A^\prime}}),
	\label{eq:nob}
\end{equation}
where $N_e$ is the total electron number in experiments, $\mathcal C^\prime$ is a parameter defined in Ref. \cite{Bjorken:2009mm} with typical value of 10,  $a_1$ and $a_2$ are fitting parameters, $\Gamma_{A^\prime}$ is the decay width of $A^\prime$, $L_{\rm sh}$ is the distance of the end of the shield and $L_{\rm dec}$ is the distance of detector from the collision point. The constraints are obtained by requiring $N(\epsilon,m_{A^\prime})$ equal to the allowed signal events, for example, 17 events for E774. We adjust the fitting parameter $a_1$ and $a_2$ to reproduce the original results from the experiments. The results are shown in Fig. \ref{fig:deamdumre}, and one can see the fits are quite well. 

Then, with the obtained $a_{1,2}$ we constrain the time-varying scenario by replacing $N(\epsilon,m_{A^\prime})$ with its time average, namely
\begin{align}
	N(\epsilon,m_0,\kappa) &=\frac{1}{t_\text{exp}} \int N(\epsilon,m_{A^\prime}(t)) dt   \\ 
	&= \frac{1}{\tau} \int_{m_0}^{\sqrt{1+\kappa}m_0} N(\epsilon,m_{A'}) \left|  \frac{d t}{d m_{A'}}  \right| d m_{A'}.\nonumber
\end{align}
Finally, setting $N(\epsilon,m_0, \kappa)$ to the allowed signal events, we obtain the limits for the time-varying scenario.

\subsection*{The Analysis of the CMS Open Data}

We test the time-varying scenario using real collider data. Unfortunately, the BaBar data are unavailable, while the LHCb and ATLAS Open Data are inadequate for scientific study. Only the CMS Open Data provides the full complexity of the collision data and is adopted in our analysis. We analyze the CMS dimuon sample in the AOD format, collected in 2012 with a collision energy of $\sqrt{s}=8$ TeV~\cite{CMSDoubleMu2012RunB, CMSDoubleMu2012RunC}. The CMS has recorded an integrated luminosity of $20.6~{\rm fb}^{-1}$ for 8 TeV in 2012~\cite{CMS:2014lcz}, but only provide $50\%$ to the public in the form of CMS Open Data~\cite{Lassila-Perini:2021xzn} according to the CMS data policy~\cite{cms2012cms}.

\begin{figure*}[hbt]
	\centering
	\includegraphics[width=0.7\textwidth]{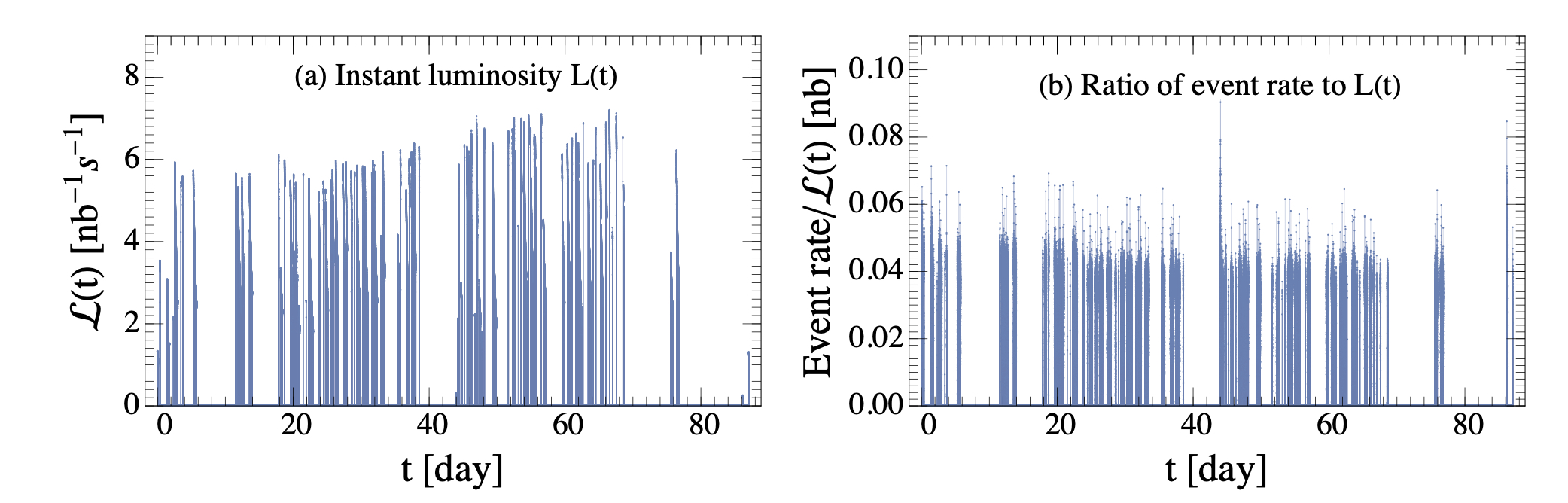}
	\caption{{\bf Luminosity records from the CMS Open Data.} (a): the instant luminosity $\mL(t)$ recorded by the CMS dimuon data 2012~\cite{CMSDoubleMu2012RunB, CMSDoubleMu2012RunC}, with time $t=0$ starting from 2012.07.07, 23:14:58. (b): the ratio between the event rate and instant luminosity is approximately constant.}
	\label{fig:lumi}
\end{figure*}

For a time-varying signal, the greatest challenge is that the actual experimental data are not taken continuously and uniformly. The data delivering follows the scheduled program, which leads to a non-uniform instant luminosity $\mL(t)$, as shown in the left panel of Fig.~\ref{fig:lumi}. 
In the right panel of Fig.~\ref{fig:lumi}, we plot the number of dimuon events in a small time interval (10 seconds) and divide it by the integrated luminosity in this interval. As expected, the event rate per unit luminosity is roughly constant since SM cross-section is constant with time.
The instant luminosity induces time dependence from human operation aside from the intrinsic time-oscillation of the signal. Therefore, we must improve our previous analysis, which assumed a uniform and continuous instant luminosity, and adapt improved strategies to the experiment data.

For the {\bf  double-peak method}, we perform a one-dimensional analysis on the dimuon invariant mass distribution $m_{\ell\ell}$ with $\ell=\mu$. In general, the distribution should be
\be\label{time-varying}
\frac{d\sigma_S}{dm_{\ell\ell}}=\epsilon_S(m_{\ell \ell}) \sigma_0(m_{\ell \ell}) \times \delta\left(m_{\ell\ell}-m_{A'}(t)\right),
\ee
where $\epsilon_S(m_{\ell \ell})$ and $\sigma_0(m_{\ell \ell})$ are the cut efficiency and  the production cross section of $pp\to A'(\to\ell^+\ell^-)X$ with $m_{A'} = m_{\ell \ell}$, $\delta(...)$ is the Dirac $\delta$ function, $m_{A'}(t)=m_0\sqrt{1+\kappa\cos^2(m_\phi t+\phi_0)}$ is the time-dependent resonant mass varying from $m_0$ to $\sqrt{1+\kappa}m_0$, and the oscillation period is $\tau=\pi/m_\phi$. The $\delta$ function can be rewritten as
\begin{align}\label{eq:delta}
&\delta\left(m_{\ell\ell}-m_{A'}(t)\right) \nonumber \\
&=\frac{2m_{\ell\ell}}{\pi\sqrt{m_{\ell\ell}^2-m_0^2}\sqrt{(1+\kappa)m_0^2-m_{\ell\ell}^2}} \nonumber\\
&~~\times\left(\frac{\pi}{2m_\phi}\right) \left[\sum_{i=-\infty}^{+\infty}\delta\left(t-t_i^+(m_{\ell\ell})\right)+\delta\left(t-t_i^-(m_{\ell\ell})\right)\right] \nonumber\\ 
&= \frac{f(y)}{m_0} \frac{\tau}{2} \left[\sum_{i=-\infty}^{+\infty}\delta\left(t-t_i^+(m_{\ell\ell})\right)+\delta\left(t-t_i^-(m_{\ell\ell})\right)\right],
\end{align}
where $y = m_{\ell \ell}/m_0$, $f(y)$ is the probability density function for the invariant mass spectrum, $t_i^\pm$ are the two solutions for $m_{A'}(t)=m_{\ell\ell}$ for the $i$th oscillation period, with $i\in\mathbb{Z}$. Their explicit expressions are
\begin{equation}
\begin{aligned}
t_i^+(m_{\ell\ell})=\frac{1}{m_\phi}\left[i\pi+\arccos\sqrt{\frac{m_{\ell\ell}^2-m_0^2}{\kappa m_0^2}}-\phi_0\right],\\
t_i^-(m_{\ell\ell})=\frac{1}{m_\phi}\left[i\pi-\arccos\sqrt{\frac{m_{\ell\ell}^2-m_0^2}{\kappa m_0^2}}-\phi_0\right].
\end{aligned}
\end{equation}

The expected event number is the time integral of the product of cross section $d\sigma_S/dm_{\ell\ell}$ and instant luminosity $\mL(t)$. Substituting \Eq{eq:delta} into \Eq{time-varying}, one gets the signal event number in an invariant mass bin during a time interval $[t_1,t_2]$
\begin{equation}
\begin{aligned}
\frac{dN_S}{dm_{\ell\ell}}&=\int_{t_1}^{t_2}dt\mL(t)\frac{d\sigma_S}{dm_{\ell\ell}}\\
&=\epsilon_S(m_{\ell \ell}) \sigma_0(m_{\ell \ell})  \frac{f(y)}{m_0}
\frac{\tau}{2}\left[\sum_{i}\mL(t_{i}^+)+\mL(t_{i}^-)\right],
\end{aligned}
\end{equation}
where $i$ is the $i$th time period within $[t_1,t_2]$. 
If the integrated luminosity is denoted as $\text{L}$, we can define the probability density function for the real data $\bar{f}(y)$ as
\begin{equation}
\begin{aligned}
\bar{f}(m_{\ell\ell}/m_0)&\equiv\frac{1}{\text{L} \epsilon_S(m_{\ell \ell}) \sigma_0(m_{\ell \ell}) }\frac{dN_S}{d(m_{\ell\ell}/m_0)}\\
&=f(y) \frac{\tau}{2 \text{L}} \left[\sum_{i}\mL(t_{i}^+)+\mL(t_{i}^-)\right].
\label{eq:varies}
\end{aligned}
\end{equation}
When the oscillation period is smaller than the typical $\mL(t)$ variation time (about a few hours), the instant luminosity in each period is roughly constant, e.g., $\mL(t) = \mL_i^0$ in the $i$th oscillation period. Therefore, for $\tau \lesssim$ a few hours, we have 
\begin{align}
	\frac{\tau}{2 } \left[\sum_{i}\mL(t_{i}^+)+\mL(t_{i}^-)\right] \approx \text{L} \quad \Longrightarrow \quad \bar{f}(y) \approx f(y).
\end{align}

\begin{figure*}[tb]
	\centering
	\includegraphics[width=1\textwidth]{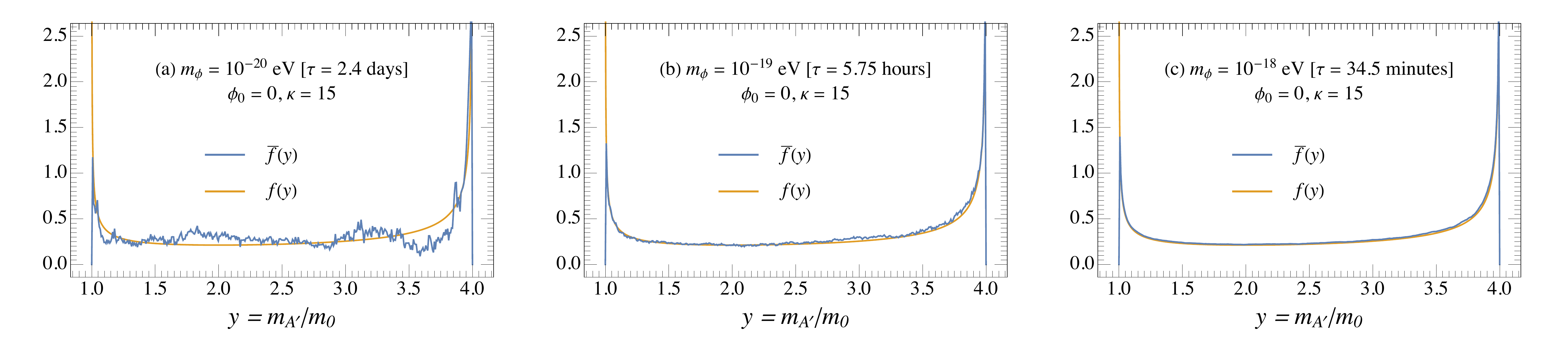}
	\caption{{\bf Realistic signal normalized spectrum based on the CMS Open Data.} The realistic spectrum $\bar{f}(y)$ is shown in blue for $ y = m_{\ell\ell}/m_0$ based on $\mL(t)$ from CMS Open Data, compared with the ideal spectrum $f(y)$ (shown in orange) assuming uniform instant luminosity. Three different oscillation periods are shown, with $m_{\phi} = 10^{-20}$ eV (a), $10^{-19}$ eV (b) and $10^{-18}$ eV (c). With small $\tau$, $\bar{f}(y)$ becomes close to $f(y)$.}
	\label{fig:realdistribution}
\end{figure*}

In Fig.~\ref{fig:realdistribution}, we use the instant luminosity $\mL(t)$ from CMS Open Data to calculate the distribution $\bar{f}(y)$ (blue) and compare it with the ideal signal distribution $f(y)$ (yellow). We choose $\kappa=15$, the initial phase $\phi_0=0$ and three different $m_{\phi}= 10^{-20},~10^{-19},~10^{-18}$ eV as benchmark points. For the blue lines of $\bar{f}(y)$, we see significant fluctuations and deviations from the yellow line $f(y)$, for $m_\phi=10^{-20}$ eV, which originates from the non-uniform instant luminosity of CMS Open Data. For $m_\phi = 10^{-19}$ eV ($\tau=5.75$ hours), the oscillation time scale is short compared with the luminosity variation time scale. Hence, the real signal spectrum $\bar{f}(y)$ is very close to the ideal spectrum $f(y)$. For even shorter oscillation period, $m_{\phi} = 10^{-18}$ eV, $\bar{f}(y)$ is almost the same as $f(y)$. Therefore, we demonstrated that for oscillation period $\tau$ smaller than a few hours, the signal invariant mass distribution is not affected by the non-uniform instant luminosity. This conclusion remains unchanged when varying the initial phase $\phi_0$.

As a result, for a small time oscillation period, e.g., $\tau \lesssim$ a few hours, the double-peak method described in the main text can apply to CMS Open Data without any further modification.
We select the 2012 dimuon sample in the CMS Open Data set and apply minimal cuts as $p_T^\mu > 15$ GeV, $|\eta| <2.4$. We aim to compare the limits for the traditional scenario $\kappa =0$ and the time-varying scenario using double-peak and time-dependent methods. In the recast for the traditional scenario , our procedure and limit using CMS Open Data are similar to the recast of Ref.~\cite{Curtin:2014cca} using CMS data in Ref.~\cite{CMS:2014lcz}.

For the DPM, we smear the signal distribution with an energy resolution $\sigma_{\rm re}=0.026~{\rm GeV}+0.013\,m_{\ell\ell}$~\cite{CMS:2011xlr}, and take bins of $m_{\ell\ell}$ with a width of $\Delta m=\sigma_{\rm re}$. For each $m_{A'}$, we fit the signal model for $m_0 = m_{A'}$ and $m_0 = m_{A'}/\sqrt{1+\kappa}$, and obtain two corresponding $\epsilon^2$ limits, as described before. The most stringent constraint for a given $m_0$, usually from the right peak, is adopted as the CMS bound for the DPM. In Fig.~5 of the main text, we stop the limit at about 10 GeV for the traditional scenario $\kappa =0$ (dashed gray) \cite{CMS:2019buh},  but the DPM analysis can extend to $m_0 = 10~{\rm GeV}/\sqrt{1+\kappa}$. 
Furthermore, we do not constrain the $Z$ pole region for the traditional scenario $\kappa =0$, resulting in a blank region. This feature is inherited by the DPM result, which appears on the left, e.g., $m_0 \approx m_Z /\sqrt{1+\kappa}$ from the right peak. However, the limit is not vanishing because this same $m_0$ can receive limits from the left-peak $m_0 = m_{A'}$. The limit is weaker but does not vanish. In summary, the DPM limit is similar to the traditional limit, shifting to the left by a factor of $\sqrt{1+\kappa}$. In addition, the DPM limit is weaker than the traditional limit comparing the corresponding points, $m_0$ for DPM and $m_0 \sqrt{1+\kappa}$ for the traditional model because the time-varying scenario has a reduced integrated luminosity compared with the traditional one. 
\\

For the {\bf time-dependent method}, we read the invariant mass $m_{\ell\ell}$ and time-stamp $t$ of each event and then fulfill the events into a two dimensional $t$-$m_{\ell\ell}$ bins. We use the letter $i$ to denote the $i$th time bin and $j$ to denote the $j$th invariant mass bin. In the 2D grid, we only take the bins along the $m_{A'}(t)$ trajectory, as illustrated in Fig.~3 of the main text, which roughly satisfies $m_j \approx m_{A'}(t_i)$. We denote these bins along the signal trajectory as $(i,j)$ pairing, where $j$ is determined by $i$.
Therefore, the expected signal event within such $(i,j)$ bin is
\begin{equation}
\begin{aligned}\label{2d}
S_{ij}&=\int_{t_i}^{t_i+\Delta t}dt\int_{m_{j}}^{m_{j}+\Delta m_{\ell\ell}}dm_{\ell\ell} \\
 &~~\times\left( \frac{1}{\sqrt{2\pi}\sigma_m}e^{-\frac{(m_{\ell\ell}-m_{A'}(t))^2}{2\sigma_m^2}}\times \mL(t) \times  \epsilon_S(m_{\ell \ell}) \sigma_0(m_{\ell \ell})\right) ,
\end{aligned}
\end{equation}
where the first factor is the Gaussian smearing of the $m_{\ell \ell}$ resolution, the second term is the instant luminosity, and the last term is signal cut efficiency times the production cross-section.
For the two-dimensional binning, we use $\Delta m=\sigma_{\rm re}$ and $\Delta t=\tau/8$. 
We have tried more fined grids for $t$, but the result has not improved. Because the current grid is already small enough, the observed event number in each grid is quite small, or even zero. Therefore, further refining the grid will suffer more from statistical error. 
Given the signal and background event numbers in each bin, we use \Eq{eq:LLR} to derive the LLR by summing up all the $(i,j)$ bins. Finally, we plot the limit from the TDM in Fig.~5 in the main text for $\kappa=15,~24$, $\phi_0 = 0$, and $m_\phi = 10^{-19}$ eV. The limit from TDM is indeed better than DPM by $1-2$ orders of magnitude.

In summary, we have successfully placed the limits using the actual collider data for a time-oscillating signal, which opens up the time-oscillating analysis to the exotic category searches at the collider. It is a pathfinder search for this broad class of time-varying signals, not limited to changing mass but also couplings, production cross-sections, and decay rates. Furthermore, we demonstrated the feasibility and the potential of a renewed class of exotic signals, namely the time oscillation signal, which is useful for experimental analysis.

\section*{Data Availability}

The data extraction code of CMS Open Data is included in \url{https://github.com/gitguojh/cms.open.data.code.git}, which is based on the example code of the dimuon spectrum from a CMS 2011 primary dataset \cite{Geiser:2017}. By analyzing the {\tt AOD} file of CMS Open Data, aside from the basic information of dimuons, such as the invariant mass $m_{\mu\mu}$, transverse momentum $p_T$, and pseudorapidity $\eta$, we also extract the the time-stamp and instant luminosity information of each event using the structure {\tt timespec} in {\tt Level1TriggerScalers.h} and the function {\tt instantLumi} in {\tt LumiScalers.h}. With this event-by-event information in hand, both the DPM and TDM can be applied. Additional details can be accessed in {\tt README.md}, {\tt demoanalyzer\_cfg.py}, and {\tt scr/DimuonSpectrum2011.cc} at the aforementioned Github link. And all relevant data are available from the corresponding author upon request.

\begin{acknowledgments}
We thank Maxim Pospelov, Wei Xue, Daneng Yang and Yanxi Zhang for valuable discussions and Bertrand Echenard for helpful discussion on BaBar's data and analysis. Furthermore, we want to thank Clemens Lange, Qiang Li, and Yiwen Wen for their help in dealing with the CMS Open Data. Finally, KPX is highly grateful for the help of programming from Guang-Ze Fu. JHG is grateful to Fanqiang Meng for his help with CMS software installation. JL is supported by the National Science Foundation of China under Grant No. 12075005, 12235001, and by Peking University under startup Grant No. 7101502458. XPW is supported by the National Science Foundation of China under Grant No. 12005009.
\end{acknowledgments}

\section*{References}

\bibliographystyle{utphys}
\bibliography{ref}
\clearpage
\newpage
\onecolumngrid
\begin{center}
	\textbf{\large Unveiling Time-Varying Signals of Ultralight Bosonic Dark Matter at Collider and Beam Dump Experiments}\\
	\textbf{\it\large Supplemental Materials} 
	
	\vspace{0.05in}
	{Jinhui Guo, Yuxuan He, Jia Liu, Xiao-Ping Wang and Ke-Pan Xie}
\end{center}


\setcounter{equation}{0}
\counterwithout{equation}{section}
\setcounter{figure}{0}
\setcounter{table}{0}
\setcounter{section}{0}
\setcounter{page}{1}
\makeatletter
\renewcommand{\theequation}{S\arabic{equation}}
\renewcommand{\thefigure}{S\arabic{figure}}
\renewcommand{\thetable}{S\arabic{table}}

\onecolumngrid

\section*{Supplementary Note 1: Comparison of the full-fitting and simplified methods at LHCb}

\begin{figure}[tbh]
	\centering
	\includegraphics[width=0.4 \textwidth]{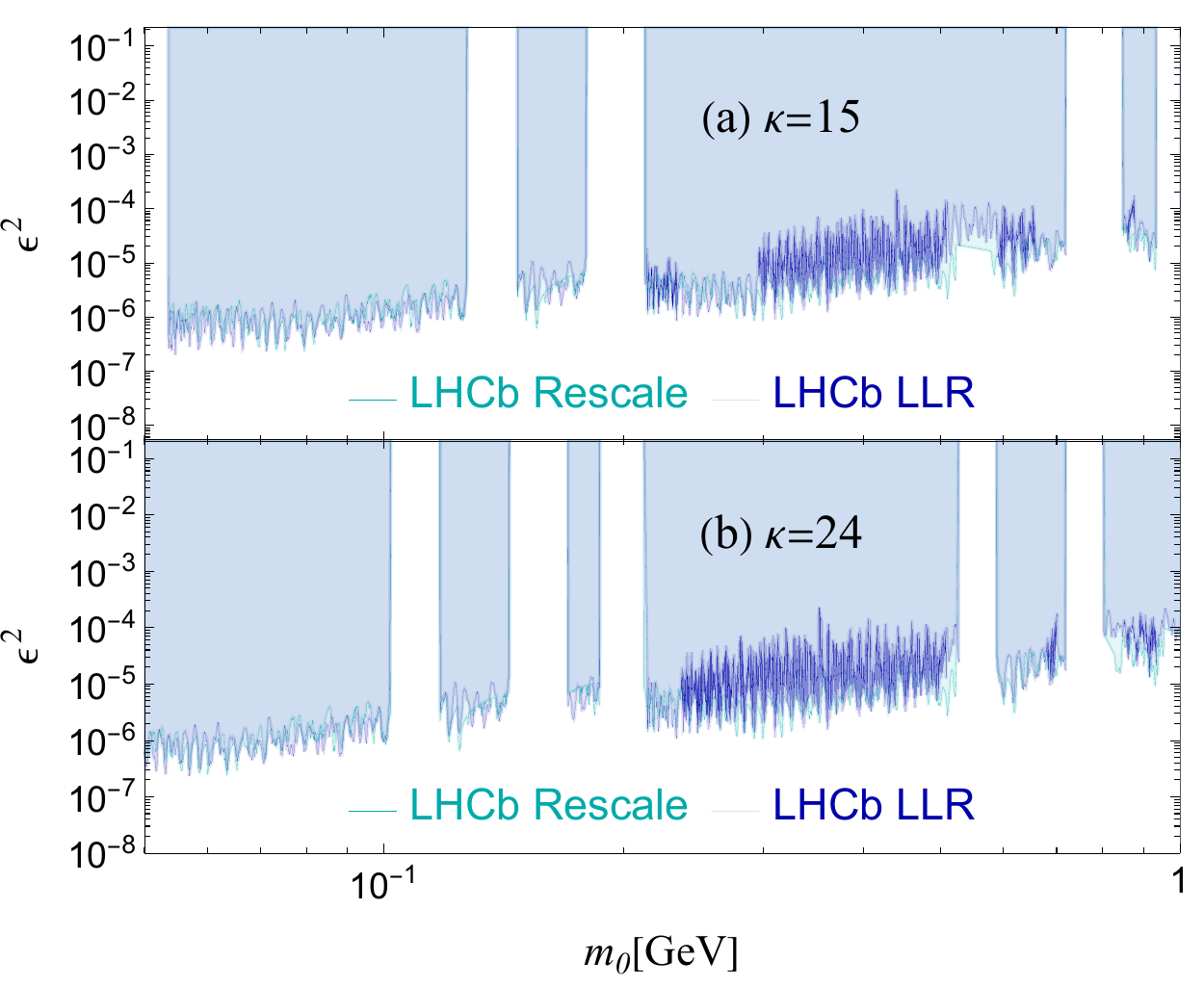}
	\caption{{\bf Recast of the LHCb data using simply rescaling and full recasting methods.} Limits on $\kappa=15$ and 24 are shown in (a) and (b), respectively. The good agreement between the two curves demonstrates the robustness of the simplified rescaling method.}
	\label{fig:lhcb-com}
\end{figure}

In the blue line of Fig.~\ref{fig:lhcb-com}, we apply the full-fitting LLR method to the $m_{\mu\mu}$ spectrum as described in the main text to recast the LHCb. The details of this application can also be found in the next paragraph of this {\it Supplemental Materials}. Besides, we apply the simplified method by rescaling the traditional $\epsilon^2$ bounds in each invariant mass bin $m_{\ell\ell}\in[m_0, \sqrt{1+\kappa}m_0]$ according to the exposure time and obtaining the most stringent one as the $\epsilon^2$ limit for a given $m_0$ in our time-varying signal model. Then we obtain the green line of Fig.~\ref{fig:lhcb-com}. The good agreement between the green and blue lines illustrates the robustness of the simplified method.

\section*{Supplementary Note 2: Prompt decay $A' \to \mu^+\mu- $ at LHCb}

\begin{figure}[htbp]
	\centering 
	\includegraphics[width= \textwidth]{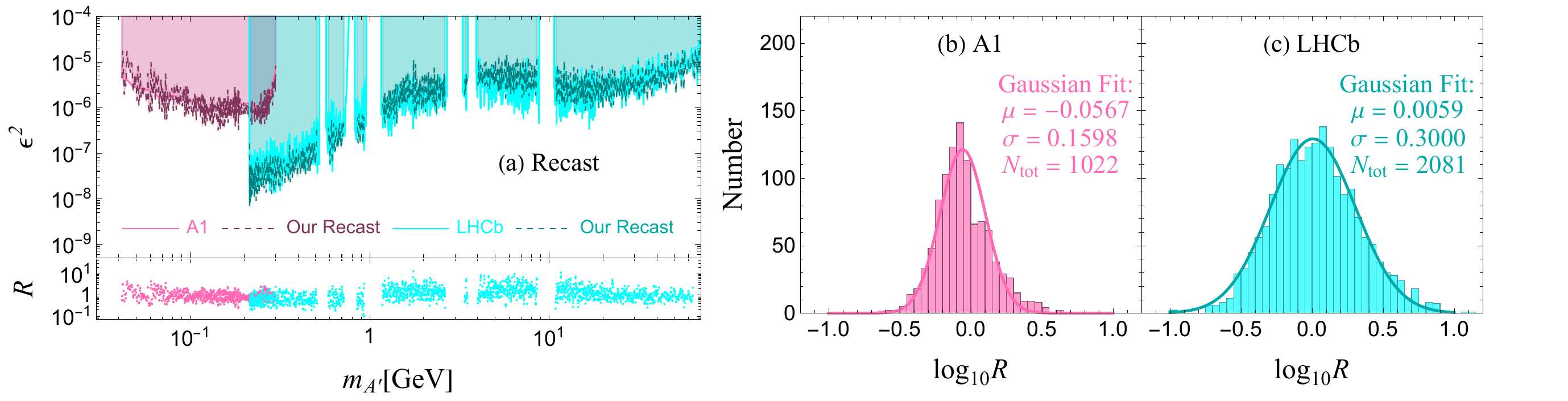}
	\caption{{\bf Recast of the A1 and LHCb data.} (a): Limits on mixing strength $\epsilon^2$ as a function of mass parameter $m_{A'}$ in A1 and LHCb experiments without the time-varying effect, where the existing and recast limits are drawn as solid and dashed respectively in the top. Moreover, the $\epsilon^2$ ratios between the recast and existing limits are plotted at the bottom.
		(b)(c): The distribution of $\log_{10}R$ for A1 and LHCb experiments.}
	\label{fig:lhcb-recast}
\end{figure}

A similar strategy can be applied to recast and reinterpret the LHCb measurements in the $pp\to A'\to\mu^+\mu^-$ channel, with an integrated luminosity of $\sim5~{\rm fb}^{-1}$. We take the $m_{\mu\mu}$ data from Ref.~\cite{LHCb:2020ysn} and use the method described in Ref.~\cite{LHCb:2019vmc} to fit the data and obtain the allowed signal event $S$.
Fortunately, the $n_{\rm ex}^{A'}[m_{A'},\epsilon^2]/\epsilon^2$ spectrum is provided in the appendix of Ref.~\cite{LHCb:2019vmc}, which contains the signal efficiency to unfold the limit on $S$ to  $\epsilon^2$.
Assuming the Gaussian smearing of the signal, we have nicely repeated the LHCb constraints as shown in Fig.~\ref{fig:lhcb-recast}(a). And in Fig.~\ref{fig:lhcb-recast}(b), the distribution of $\log_{10}R$ is shown and fitted to the Gaussian shape with $\mu \approx 0$ and standard deviation  $\sigma \approx 0.3$. Therefore, our recast is in good agreement with the LHCb official result and has a deviation of $10^{0.3} \approx 2$ at $1\sigma$ level. 

To constrain the time-varying scenario, again, we replace the template $f_G$ with $f_S$ to obtain the bounds. Similar to the BaBar experiment, for each $m_{A'}$, we perform two fits for the double peaks, respectively. In the main text, they are translated to $\epsilon^2$ limits as a function of $m_0$ for the LHCb experiment for the time-varying scenario, shown in Fig.~2.

\section*{Supplementary Note 3: Long-lived decay $A' \to \mu^+\mu^- $ at LHCb}

The LHCb collaboration also searches for the long-lived $A' \to \mu^+ \mu^-$ decay in Ref.~\cite{LHCb:2019vmc}. Here we also apply the search to the time-varying scenario.
In the LHCb analysis, the limit (90\% C.L.) on observed signal events $n_{\rm ob}^{A^\prime}[m_{A^\prime}, \epsilon^2]$ is obtained by subtracting the background model from the observed data, after considering the signal pattern. Different from prompt decay, the limits on the long-lived dark photon are derived via a two-dimensional fit relying on both the dark photon mass $m_{A^\prime}$, which determines the invariant mass of dimuon, and the kinetic mixing parameter $\epsilon$ which determines the lifetime of $A^\prime$ and the displacement of the vertex. On the other hand, the expected signal event number $n_{\rm ex}^{A^\prime}[m_{A^\prime},\epsilon^2]$ is calculated by the simulation of the signal. The LHCb constraint is obtained by requiring $n_{\rm ob}^{A^\prime}[m_{A^\prime},\epsilon^2] \leqslant n_{\rm ex}^{A^\prime}[m_{A^\prime},\epsilon^2]$, which sets the limit on kinetic mixing $\epsilon^2$ for a given mass $m_{A^\prime}$. 

\begin{figure}[htb]
	\centering
	\includegraphics[width=0.35 \linewidth]{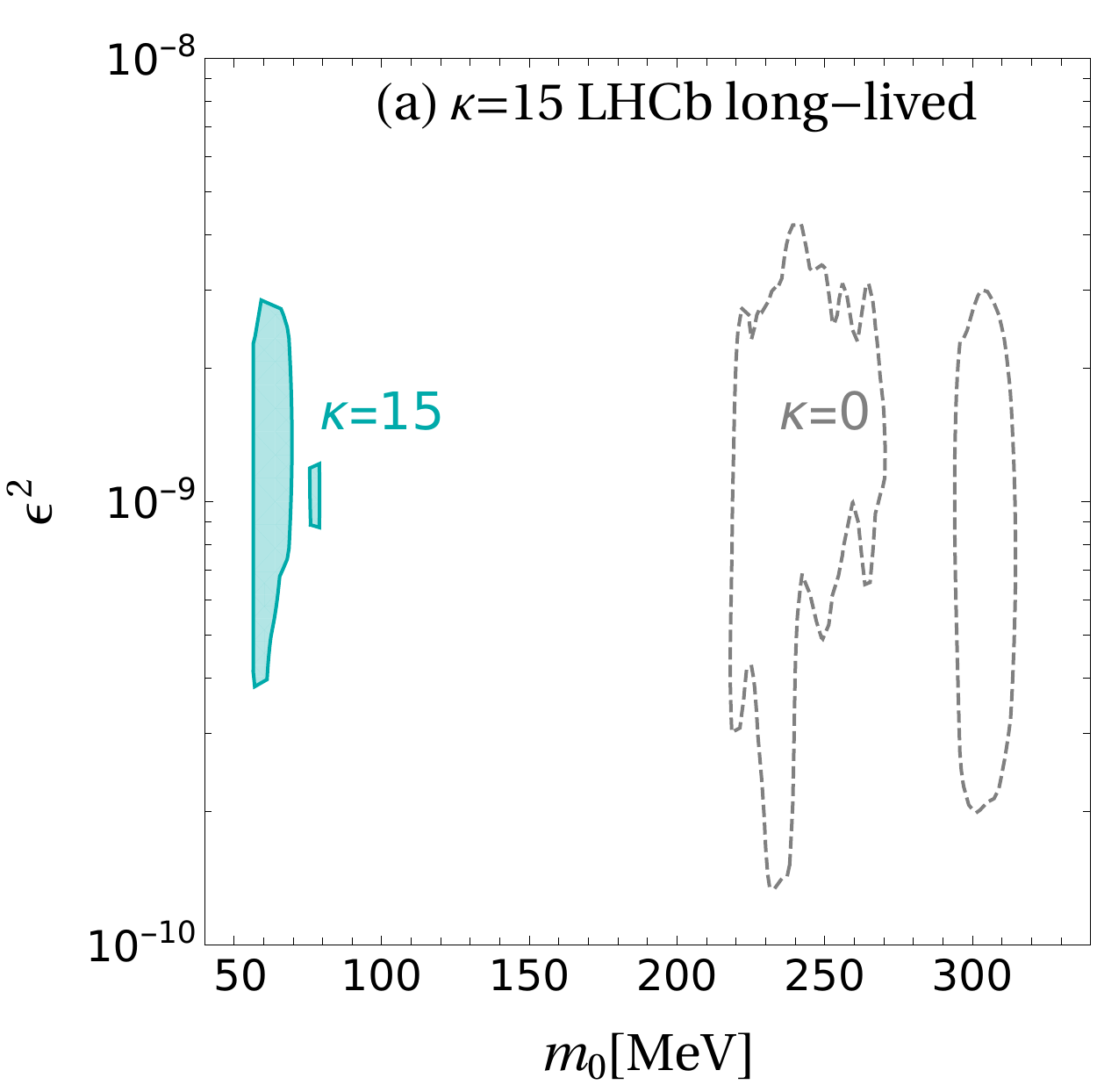}\qquad
	\includegraphics[width=0.35 \linewidth]{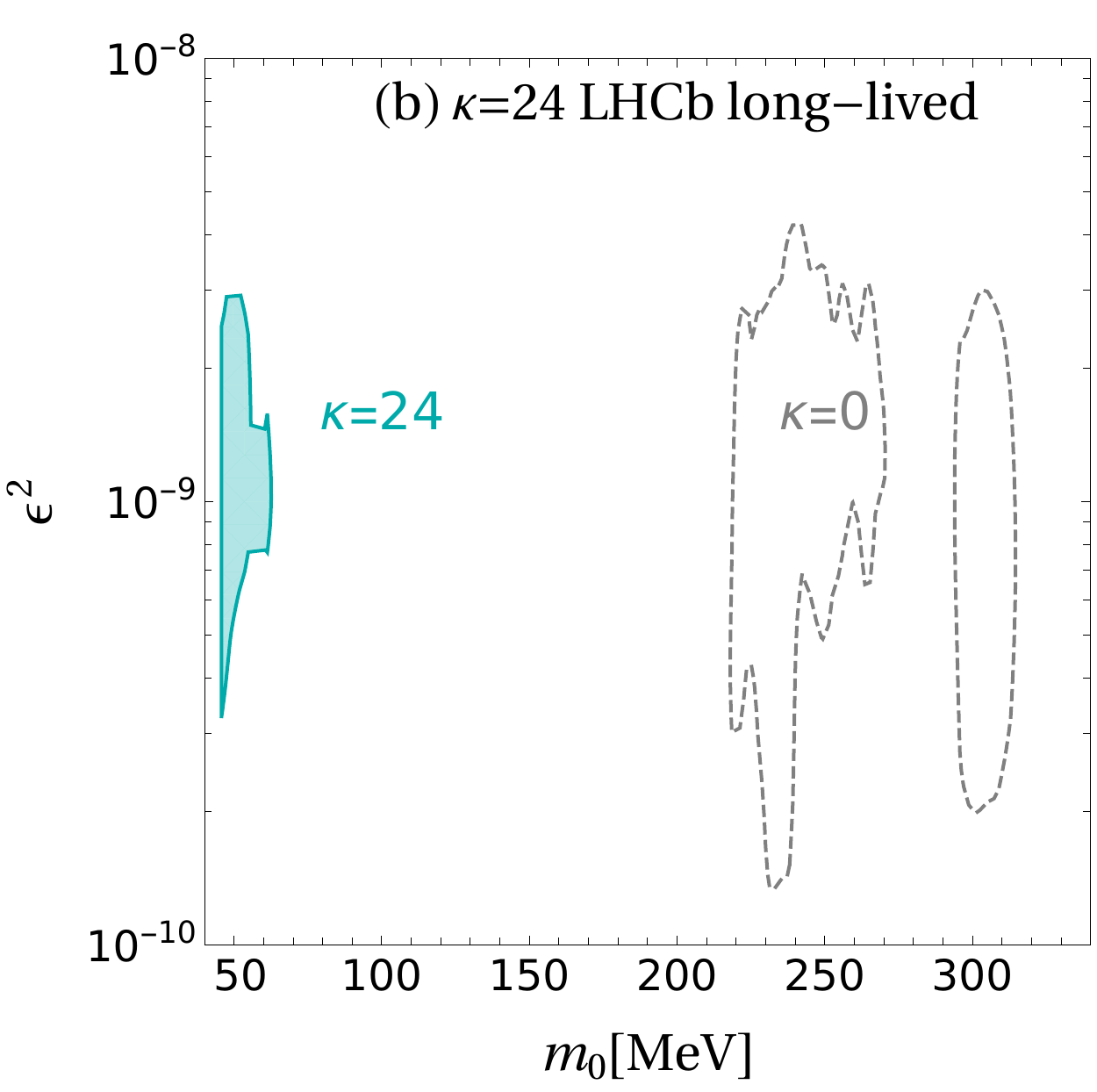}
	\caption{{\bf Excluded parameter regions from LHCb long-lived search.} The time-varying scenario $\kappa=15$ and 24, are shown in the cyan regions in (a) and (b), respectively. The official excluded regions, for the traditional resonant scenario $\kappa = 0$, are shown in dashed contours.}
	\label{fig:lhcbll}
\end{figure}

Regarding the time-varying scenario, we obtain the constraints by averaging the spread of the signal events due to the time-varying effect. In our scenario, the dark photon mass is a function of time, $m_{A'}(t)$, and ranges from $m_0$ to $\sqrt{1+\kappa}m_0$. The probability of $m_{A'}$ taking a particular resonant mass $m_{\rm res}$ is described by the probability density function $f(y)$. Therefore, in the time-varying scenario, the observed limit $\bar{n}_{\rm {ob}}^{A^\prime}[m_0,\kappa,\epsilon^2]$ and the expected number of events $\bar{n}_{\rm {ex}}^{A^\prime}[m_0,\kappa,\epsilon^2]$ becomes
\begin{equation}
	\bar{n}_{\rm {ob/ex}}^{A^\prime}[m_0,\kappa,\epsilon^2] = \sum_{i} n_{\rm {ob/ex}}^{A^\prime}[m_{i},\epsilon^2] \times \int_{m_i}^{m_{i+1}} \frac{d \bar{m}}{m_0} f\left(\frac{\bar{m}}{m_0}\right),
	\label{eq:nAVE}
\end{equation}
where $i$ is the $i$th bin for the invariant mass spectrum, $m_i$ is the mass value for the $i$th bin. In Eq.~(\ref{eq:nAVE}), the last factor $\int_{m_i}^{m_{i+1}} \frac{d \bar{m}}{m_0} f\left(\frac{\bar{m}}{m_0}\right)$ describes the probability of $m_{A'}$ falls into the mass bin $\left[m_i,~ m_{i+1} \right]$. Therefore, the event numbers $\bar{n}_{\rm {ob/ex}}^{A^\prime}[m_0,\kappa,\epsilon^2]$ in the time-varying scenario are the average of $n_{\rm ob/ ex}^{A^\prime}[m_{A^\prime},\epsilon^2]$ over the random variable $m_{A'}$. The limits on the time-varying scenario can be similarly obtained by setting $\bar{n}_{\rm {ob}}^{A^\prime}[m_0,\kappa,\epsilon^2] \leq \bar{n}_{\rm {ex}}^{A^\prime}[m_0,\kappa,\epsilon^2]$. 

In Fig.~\ref{fig:lhcbll}, we plot the excluded regions for $\kappa = 15,~24$ of the time-varying scenario in the cyan-shaded region and compare it with the traditional resonance model ($\kappa =0$) in gray dashed contours. First, the exclusion regions are much smaller than $\kappa =0$. Second, the exclusion region shift to the left by a factor around $\sqrt{1+\kappa}$. These two features can be understood from the double-peak feature of the signal. The figure shows the sensitivity region of LHCb in the gray dashed contours for traditional resonance scenario, around $m_A'=m_{\mu^+\mu^-} \sim 250 $ MeV and $\epsilon^2 \sim 10^{-9}$.  A time-varying signal has a double-peak feature in the invariant mass spectrum, one is $m_0=m_A'$, and the other is $\sqrt{1+\kappa}m_0=m_A'$. The right peak of $m_{A'}$ has a probability larger than the left peak by $\sim (1+\kappa)^{1/4}$, as shown in Eq.~(13) in the main text. Therefore, the exclusion region of the time-varying scenario shrinks and shifts to the left around $m_0\sim 250/\sqrt{1+\kappa}$ MeV.

\section*{Supplementary Note 4: $A' \to e^+e^- $ decay at A1}
 
An analogous LLR calculation is performed to A1 experiment in the fixed-target electron scattering process $e^-Z\to e^-ZA'$, with prompt decay $A'\to e^+ e^- $ for $0.040~\text{GeV}\lesssim m_{A'}\lesssim 0.300~\text{GeV}$ in 2014 \cite{Merkel:2014avp}.  With the data taken from Ref. \cite{Merkel:2014avp}, we have recasted the exclusion limits with Gaussian smearing. In the left panel of Fig.~\ref{fig:lhcb-recast}, one can see the recast is very close to the A1 official result for the fixed resonant scenario. The A1 experiment~\cite{Merkel:2014avp} provides $m_{e^+ e^-}$ data down to 30 MeV. However, the number of events there is quite small and thus will suffer from large statistical errors. Therefore, their official analysis stops at 40 MeV. 
In addition, we plot the ratio $R$ between our recast and the A1 official result and the distribution of $\log_{10}R$ in Fig.~\ref{fig:lhcb-recast}. The results show that our recast agrees with the official results well. The deviation is tiny, within a factor of $10^{0.16} = 1.45$ at 1$\sigma$ level. Like BaBar and LHCb experiment, the double-peak method is again applied to obtain new limits on $\epsilon^2$ as a function of $m_0$.



\end{document}